\documentclass[12pt,a4paper,twoside]{article}
\usepackage[latin9]{inputenc}
\usepackage{esint}

\def\be{\begin{equation}}
\def\ee{\end{equation}}
\def\bea{\begin{eqnarray}}
\def\eea{\end{eqnarray}}

\def\6{\partial} \def\al{\alpha} \def\b{\beta}
\def\g{\gamma} \def\d{\delta} \def\ve{\varepsilon}
\def\e{\epsilon} 
 \def\h{\eta} \def\th{\theta}
 \def\k{\kappa} 
\def\m{\mu}   \def\p{\pi}
 \def\s{\sigma} \def\t{\tau}
\def\p{\phi}  \def\ps{\psi}
\def\o{\omega}  
  
  \def\O{\Omega}
 \def\la{\large} 
  
  \def\a{\dag}
\def\non{\nonumber\\}
\def\ve{\vec}

\def\v{\upsilon}
  
\def\le{\left}
\def\ri{\right}
\def\fr{\frac}
\def\bo{\boldmath}
\def\n{\nabla}
\def\la{\langle}
\def\ra{\rangle}

\def\w{\widetilde}
\def\g{\gamma}

\newcommand{\ZZ}{{\cal Z}}

\newcommand{\fourmat}[4]{\left(\begin{array}{cc}  
{#1} & {#2} \\ {#3} & {#4} \end{array} \right)}

\title{Thermo Field Dynamics of strings with definite boundary conditions}

\author{Ion V. Vancea \\
    {\em Grupo de F\'{\i}sica Te\'{o}rica e Matem\'{a}tica F\'{\i}sica,}\\
    {\em Departamento de F\'{\i}sica,}\\
        {\em Universidade Federal Rural do Rio de Janeiro (UFRRJ)}\\
    {\em Cx. Postal 23851, BR 465 Km 7, 23890-000 Serop\'{e}dica - RJ, Brasil}
    }

\date{24 August 2015} 

\makeatother

\begin{document}
\maketitle 
\begin{abstract}
In this paper we review the construction of the thermal bosonic string and $D$-brane in
the framework of the Thermo Field Dynamics (TFD). We briefly recall the well-known 
light-cone quantization of the bosonic string in the conformal gauge in flat space-time.
Then we give a bird's eye view of the fundamental concepts of the TFD. Also, we present the thermalization
of the bosonic string and the construction of the thermal $D$-brane boundary state. Finally, we 
show the calculation of the entropy of the thermal open string states with all 
boundary conditions and the entropy of the thermal $D$-brane state.
\end{abstract}

\tableofcontents{}

\section{Introduction}

At the first look, the String Theory seems just an interesting and
non-trivial application of the Quantum Mechanics and the Special Relativity
to vibrating strings. By itself, the quantization of relativistic
strings does not call the attention of the particle physicist as a
significant paradigm shift. However, when the string quantization
is performed by applying the standard rules of the perturbative Quantum
Field Theory, one discovers that the strings in certain states have
the same physical properties as the gravity in the flat space-time.
Moreover, when the condition that the space-time be flat is relaxed,
it follows that the consistency of the string degrees of freedom
with the rules of the Quantum Mechanics (the algebraic structure of
the string operators) is satisfied only if the space-time metric obeys
the Einstein's equations for the gravitational field 
\begin{equation}
l_{s}^{2}R_{\mu\nu}+l_{s}^{4}R_{\mu\rho\sigma\chi}R_{\nu}^{\rho\sigma\chi}+\cdots=0,\label{rev-Einstein-eq}
\end{equation}
where $l_{s}$ is the string length. These two facts are remarkable
because there is no a priori connection between the quantization of
relativistic strings and gravity, yet the gravitational interaction
shows up at various levels during this process. The String Theory is, at the present time, 
the unique theory in which the gravity emerges
formally from the simultaneous application of the Quantum Mechanics
and the Special Relativity together with the rest of fundamental interactions.
While that has been the main driving motivation for the study of the String Theory 
during the last four decades or
so, several other deep results have continuously emerged from String
Theory that concern the dual interpretation of the quantum matter and the
space-time in various mathematical models of the fundamental interactions.
The complexity of the mathematical tools needed to formulate the deepest
physical constructs represents a major difficulty in making fast progress
in understanding the String Theory. We are still far from understanding
the new aspects of the reality proposed by the String Theory because
they form a coherent system of concepts only in space-time manifolds
with higher dimensions than we have been able to observe. These manifolds 
represent a vast multiplicity of possible non-equivalent universes.
Another difficulty encountered in the development of the String Theory
is the lack of experimental guidance. This is due to the fact that
the typical scale of processes described by the 
string interactions is far beyond
our present and near future experimental possibilities. Nevertheless,
since the inception of the String Theory, a serious effort has been
put into formulating and deriving the fundamental physical phenomena
in the string framework such as the calculation of phenomenological
Particle Physics processes and data, the description of Statistical
and thermodynamical systems and processes, the development
of Cosmological concepts, etc.

In the present paper I will review the basic concepts of a set of fundamental results
aimed at understanding the microscopic structure of the thermal properties of strings and $D$-branes, 
their natural higher dimensional generalization. This line of research has as an objective the development of the fundamental tools necessary to formulate the dynamics of strings at finite temperature. However, it is equally important for modelling concrete problems in the framework of the String Theory such as the description of the primordial universe in which the high temperatures and energy densities constituted natural physical conditions.
The implementation of temperature in the String Theory can be performed in several formalism which at equilibrium are 
are shown to be equivalent for different systems. In this paper I will restrict the presentation to the Thermo Field Dynamics (TFD) method which is an operatorial approach to the quantum field theory at finite temperature. The main advantage of the TFD against other formalisms is that it maintains the structure of the field excitation explicit. This is useful for studying the thermodynamical properties of the perturbative $D$-branes which are described by boundary states constructed out of states from the Fock space of string (but not a proper string state). Also, the TFD gives a clear formulation of the temperature induced symmetry breaking and since the time variable is not compactified, it useful to investigate the time evolution of the thermal fields.  

The TFD method was proposed for the first time by Takahashi and Umezawa in \cite{ut}. It is a quantum field theory at finite temperature constructed in the canonical quantization that is known to be equivalent at thermodynamical equilibrium with the path integral formulation (see, e. g. \cite{hu,Khanna:2009zz}). The TFD was applied to the String theory in the seminal paper by Leblanc in \cite{Leblanc:1987hw}. Other early applications of TFD to string models were concerned with the study of 
the cosmological constant \cite{Leblanc:1988eq,Fujisaki:1997sf}, 
the regularization and the renormalization of the string propagators at finite temperature and of the free energy 
\cite{Leblanc:1989gs,Fujisaki:1989kp,Leblanc:1989yy,Fujisaki:1989sy,Fujisaki:1989ss,Leblanc:1990ub,Fujisaki:1991as,Nakagawa:1993kr,Fujisaki:1993bx,Fujisaki:1992wb}, 
the phase structure and the thermal duality
\cite{Fujisaki:1996vz,Fujisaki:1997rc,Fujisaki:1996nx},
the TFD of the heterotic string
\cite{Fujisaki:1997en,Fujisaki:1997sp}
and the dynamical mass generation 
\cite{Clavelli:1991at}.
The interest in the TFD formulation of strings at finite temperature was revived after the discovery of the $D$-brane states by Polchinski \cite{pol}. Since these are physical objects, it seems natural to try to find out their microscopic structure and physical properties at finite temperature. The first formulation of the thermal bosonic $D$-brane states was given in \cite{Vancea:2000gr} and it was extended to other string models in 
\cite{Abdalla:2001ad,Abdalla:2002ak,Abdalla:2002sb,Abdalla:2003ki,Abdalla:2003dx,Abdalla:2004xs,Vancea:2006wf,Vancea:2006tt,Vancea:2007ux,Nardi:2010xc}. The recent increasing interest in solving the String Theory in various space-time backgrounds has led to the investigation of their thermal properties in the TFD framework in 
\cite{Abdalla:2000bg,Abdalla:2004dg,Nedel:2004gy,Abdalla:2004hx,Abdalla:2004xg,Abdalla:2004xn,Graca:2005sk,Belich:2006wc}.
Also, there are few studies concerned with the treatment of the String Field Theory beside the work of Leblanc \cite{Leblanc:1987zj}( see for more recent results \cite{Abdalla:2005qs,Arafeva:2015},Cantcheff:2015fia). Very recently, the TFD has been used to formulate the dynamics of strings in a time-dependent background \cite{Nardi:2011sk}.

In the following, we are going to review the application of the TFD to the bosonic open and closed strings. In the Section 2 we will present the basic relations that underlie the string dynamics and its quantization in the flat Minkowski space-time and in the conformal gauge on the world-sheet and in the light-cone gauge in the space-time. The results from this section are standard and can be found in many textbooks and lecture notes. We have freely followed the the books \cite{Green:1987sp,Polchinski:1998rq} for references on string theory and \cite{Johnson:2005mqa,Szabo:2004uy,DiVecchia:1999rh,Vancea:2001hj} for $D$-branes. In the Section 3 we will review the TFD formalism and exemplify it in the case of the non-relativistic scalar field. In the Section 4, we will show how the TFD method can be used to construct the thermal strings and $D$-brane states and to calculate their energy. We conclude in the last section.

\section{Strings at zero temperature}

In this section we review the basic results from the String Theory that are necessary to develop the formalism of
the thermal strings. The material presented in this section is standard. We will follow mainly \cite{Green:1987sp} and \cite{Polchinski:1998rq}.  

\subsection{Classical strings}

The fundamental strings are one dimensional objects of typical length
$l_{s}\sim10^{-18}GeV$. This property implies that the string phenomenology
is beyond the current and foreseeable experimental reach. Also, due
to its energy scale $m_{s}=l_{s}^{-1}$, the string must be an object
with quantum or generalized quantum properties.

The string degrees of freedom are given in terms of embeddings $X^{\mu}$
from a two dimensional manifold $\Sigma^{2}$ (the string world-sheet)
to another manifold $M^{d}$ (the space-time). The coordinates on
the world-sheet are denoted by $\sigma^{\alpha}=(\sigma^{0},\sigma^{1})=(\tau,\sigma)$.
Here, $\sigma^{0}=\tau\in I\subset\mathbf{R}$ is a time-like coordinate
necessary to describe the time evolution of the string and $\sigma^{1}=\sigma\in[0,2\pi]$
is space-like and localizes points on the string. The two possible
topologies of strings are open and closed. The fields $X^{\mu}(\tau,\sigma)$
describe the dynamics of the bosonic string in space-time. Fermionic
fields can be added to the theory. The most elegant and consistent
way to do that is by requiring either a two dimensional supersymmetry
on $\Sigma^{2}$ (the Ramond-Neveu-Schwarz formalism) or a higher
dimensional supersymmetry in the space-time $M^{d}$ (the Green-Schwarz
formalism). Although the mathematical structure of the two supersymmetric
formulations is different, it is shown that the physical degrees of
freedom and their dynamics coincide in the Minkowski space-time $\mathbf{R}^{1,9}$
\cite{Green:1987sp}. A third mathematical formalism for the supersymmetric
strings in terms pure-spinors can be constructed in ten dimensions. It
has the advantage of preserving the covariance in the quantization
process \cite{Berkovits:2000nn}.

\subsubsection{String action}

The dynamics of the bosonic string can be derived from the Polyakov
action 
\begin{equation}
S[X,h]=-\frac{1}{4\pi\alpha'}\int d^{2}\sigma\sqrt{-h}h^{\alpha\beta}(\sigma)g_{\mu\nu}(X)\partial_{\alpha}X^{\mu}\partial_{\beta}X^{\nu}.\label{rev-Polyakov}
\end{equation}
Here, the parameter $\alpha'=l_{s}^{2}$ is inverse to the string
tension, $h_{\alpha\beta}(\sigma)$ is the world-sheet metric,
$h=\det h_{\alpha\beta}$ and $g_{\mu\nu}(X)$ is the space-time metric.
The Polyakov action is covariant simultaneously in two and $d$-dimensions.
Beside the string fields $X^{\mu}(\sigma)$, it contains new dynamical
fields which are the components of the two-dimensional metric $h_{\alpha\beta}(\sigma)$.
Therefore, there are gauge symmetries of the action associated to
the transformations of these fields locally on the world-sheet.

The action (\ref{rev-Polyakov}) is the most general action that can
be written for a string that propagates in an arbitrary space-time.
However, since the main proposal of the String Theory is that the
strings are the {\em fundamental objects} in Nature, it is crucial
to construct a quantum theory of strings. The quantization of the
action (\ref{rev-Polyakov}) by applying the standard methods is in
general impossible since one cannot define an asymptotic Hilbert space
for the string fields $X^{\mu}(\sigma)$. Indeed, these fields interact
on the world-sheet with the coupling constant $g_{\mu\nu}(X)$ and
no free theory can be defined in a consistent manner. However, it
is possible to quantize the action (\ref{rev-Polyakov}) in those
backgrounds that satisfy the equation (\ref{rev-Einstein-eq}), in
particular, in the Minkowski space-time $\mathbf{R}^{1,d-1}$. The
reason for that is that the solutions to the equation (\ref{rev-Einstein-eq})
represent backgrounds in which the string theory is conformally invariant
on the world-sheet and the fields $X^{\mu}(\sigma)$ are free
\cite{Green:1987sp}. There are other space-times in which it is possible
to obtain a quantum theory of strings, such as the Anti-de Sitter
space, the $pp$-wave background, etc. 

\subsubsection{Symmetries}

The Polyakov action in the Minkowski space-time is 
\begin{equation}
S[X,h]=-\frac{1}{4\pi\alpha'}\int d^{2}\sigma\sqrt{-h}h^{\alpha\beta}(\sigma)\eta_{\mu\nu}\partial_{\alpha}X^{\mu}\partial_{\beta}X^{\nu}.\label{rev-Polyakov-flat}
\end{equation}
The action (\ref{rev-Polyakov-flat}) has global Poincaré symmetries
\begin{eqnarray}
\delta X^{\mu} & = & {\Lambda}_{\nu}^{\mu}X^{\nu}+a^{\mu},\label{rev-Poincare-X}\\
\delta h_{\alpha\beta} & = & 0,\label{rev-Poincare-h}
\end{eqnarray}
where ${\Lambda}_{\nu}^{\mu}$ is a Lorentz group element in the fundamental
representation and $a^{\mu}$ is a constant vector field. The corresponding
two-dimensional Noether currents are associated to the linear momenta
\begin{eqnarray}
j_{\alpha}^{\mu\nu} & = & -\frac{1}{4\pi\alpha'}\left(X^{\mu}\partial_{\alpha}X^{\nu}-X^{\nu}\partial_{\alpha}X^{\mu}\right),\label{rev-charge-Lorentz}\\
j_{\alpha}^{\mu} & = & -\frac{1}{2\pi\alpha'}\partial_{\alpha}X^{\mu}.\label{rev-charge-translation}
\end{eqnarray}
The currents are conserved on-shell $\partial^{\alpha}j_{\alpha}^{\mu}=0.$
The Noether charges generate the Lorentz transformations and translations
in space-time, respectively.

The gauge transformations are due to the freedom to reparametrize
and rescale the world-sheet metric. The reparametrization transformations
correspond to the change in the world-sheet coordinates $\sigma^{\alpha}\rightarrow\sigma^{'\alpha}=f^{\alpha}(\sigma)$
with $f^{\alpha}$ arbitrary smooth functions on $\Sigma^{2}$. Under
infinitesimal transformations $f^{\alpha}(\sigma)=\zeta^{\alpha}(\sigma)$,
the fields change as 
\begin{eqnarray}
\delta X^{\mu} & = & \zeta^{\alpha}{\partial}_{\alpha}X^{\mu},\label{rev-reparam-X}\\
\delta h_{\alpha\beta} & = & \zeta^{\rho}{\partial}_{\rho}h_{\alpha\beta}+2h_{\rho(\beta}{\partial}_{\alpha)}\zeta^{\rho}.\label{rev-reparam-h}
\end{eqnarray}
Under the rescaling of the metric (the {\em Weyl transformations}) the fields
change as 
\begin{eqnarray}
\delta X^{\mu} & = & 0,\label{rev-Weyl-X}\\
\delta h_{\alpha\beta} & = & \Lambda(\sigma)h_{\alpha\beta},\label{rev-Weyl-h}
\end{eqnarray}
where $\Lambda(\sigma)$ is an arbitrary smooth function. As a consequence
of the Weyl symmetry the trace of the energy-momentum tensor vanishes
$T_{\alpha}^{\alpha}=0$, where 
\begin{eqnarray}
T_{\alpha\beta}=-\frac{2\pi\alpha'}{\sqrt{-h}}\frac{\delta S[X,h]}{\delta{h}^{\alpha\beta}}.\label{rev-eq-mot-h}
\end{eqnarray}
The relation (\ref{rev-eq-mot-h}) contains the equations of motion
of the metric components. The equations of motion of $X^{\mu}$ are
\begin{eqnarray}
\partial_{\alpha}\left(\sqrt{-h}h^{\alpha\beta}\partial_{\beta}X^{\mu}\right)=0.\label{rev-eq-mot-X}
\end{eqnarray}
The above equations describe the dynamics of the classical strings
on an arbitrary two dimensional manifold. However, the existence of
the gauge symmetries given by the equations (\ref{rev-reparam-X}) - (\ref{rev-Weyl-h})
shows that the theory is formulated in terms of redundant degrees
of freedom. One has to reduce the number of string fields to the physical
degrees of freedom before proceeding with the study of the strings,
either at the classical or at the quantum level. The classical dynamics
is determined completely by the classical equations of motion (\ref{rev-eq-mot-X})
that can be simplified by using the re-parametrization and the Weyl
symmetries.

In order to quantize the string, one has to fix the gauge symmetries.
Since the local symmetries act on the metric, the gauge fixing amounts
to fixing the form of $h_{\alpha\beta}$. There are three independent
components of $h_{\alpha\beta}$. The reparametrization symmetry
can be used to fix two components and the Weyl symmetry can be used
to fixed the third one. Using these transformation, one can put the
metric in the flat form with the Minkowski signature 
\begin{equation}
h_{\alpha\beta}=\eta_{\alpha\beta}.\label{re-flat-ws-metric}
\end{equation}
The world-sheet metric is locally conformally flat since there are conformal
symmetries left on $\Sigma^{2}$ which make the quantum theory solvable.

\subsubsection{Equations of motion and constraints}

As we have concluded from the previous section, the starting point in the study of the bosonic string theory
is the conformally invariant action 
\begin{equation}
S=-\frac{1}{4\pi\alpha'}\int d^2 \sigma \partial^{\alpha} X^{\mu} \partial_{\alpha} X_{\mu}.
\label{rev-Polyakov-conformal}
\end{equation}
The action (\ref{rev-Polyakov-conformal}) is Poincaré invariant in
the flat space-time $\mathbf{R}^{1,d-1}$. It is convenient to use the notation $\sigma^{\alpha} = (\tau,\sigma)$. 
The string fields $X^{\mu}(\tau,\sigma)$ satisfy the massless Klein-Gordon equation in two dimensions 
\begin{equation}
\left( \partial_{\tau}^{2} - \partial_{\sigma}^{2} \right) X=0.
\label{rev-eq-X}
\end{equation}
In order to describe the dynamics of strings, we must consider the equations of motion of the components of the two-dimensional metric. Their direct consequence is the vanishing of the energy-momentum tensor (\ref{rev-eq-mot-h}) that takes the following form in the conformal gauge
\begin{equation}
T_{\alpha\beta} = \partial_{\alpha}X^{\mu}\partial_{\beta}X_{\mu} - \eta_{\alpha\beta} \partial^{\gamma}X^{\mu} 
\partial_{\gamma} X_{\mu}.
\label{rev-energ-mom-conf}
\end{equation}
The vanishing of $T_{\alpha\beta}$ is equivalent to the following equations
\begin{eqnarray}
\partial_{\tau}X^{\mu} \partial_{\sigma} X_{\mu} & = & 0,
\label{rev-constraint-1}\\
\left( \partial_{\tau} X \right)^2 + \left( \partial_{\sigma} X \right)^2 & = & 0.
\label{rev-constraint-2}
\end{eqnarray}
The above system of equations represent constraints among the fields $X^{\mu}$. They reduce the configuration space of the classical theory to the physical configuration space. Upon quantization, they act as constraints on the quantum string operators. The different ways to deal with the constraints differentiate the quantization methods from each other. 

\subsubsection{Boundary conditions and solutions of the equations of motion}

In order to determine the solutions of the equations of motion (\ref{rev-eq-X}), one has to provide appropriate boundary conditions. There are three types of boundary conditions that can be chosen on the bosonic string world-sheet: closed string, Neumann and Dirichlet boundary conditions, respectively. It is possible, in principle, to study the string dynamics on the euclidean two dimensional world-sheets with Robin boundary conditions, but the physical interpretation of this model is not very clear. 

The world-sheet of the closed string is homotopic with the cylinder $\Sigma^2 \sim I \times S^1 $ where $I$ is a real interval $I \subseteq \mathbf{R}$. The closed string boundary conditions state that the string fields are single valued along the space-like direction of the world-sheet
\begin{equation}
X^{\mu} (\tau, \sigma ) = X^{\mu} ( \tau, \sigma + 2 \pi).
\label{rev-closed-bc}
\end{equation}
The solution to the equation of motion (\ref{rev-eq-X}) with the boundary conditions (\ref{rev-closed-bc}) has the following Fourier expansion
\begin{equation}
X^{\mu} (\tau,\sigma) = x^{\mu} + {\alpha}'p^{\mu}\tau + 
i\sqrt{\frac{\alpha'}{2}}\sum_{n \neq 0}{\frac{1}{n}
\left( \alpha_{n}^{\mu}e^{-in(\tau - \sigma)} +
\beta_{n}^{\mu}e^{-in (\tau + \sigma)} \right)}.
\label{rev-sol-cs}
\end{equation}
The string fields  $X^{\mu}(\tau,\sigma)$ can be decomposed in to the left- and right-moving modes labelled by $n$. The first two terms from the right hand side of the equation (\ref{rev-sol-cs}) correspond to $n=0$ and are interpreted as the contribution of the string center of mass to the solution $X^{\mu}(\tau,\sigma)$. The constant vectors $x^{\mu}$ and $p^{\mu}$ are the coordinates and the momenta of the center of mass, respectively.

One can impose two types of boundary conditions at each end of the open string, the Neumann and the Dirichlet boundary conditions, respectively, 
\begin{eqnarray}
\mathrm{Neumann:} \hspace{0.5cm}  \partial_{\sigma}X^{\mu}(\tau,0) & = & \partial_{\sigma}X^{\mu}(\tau,\pi) = 0, 
\label{rev-Neumann-bc}\\
\mathrm{Dirichlet:} \hspace{0.5cm} \partial_{\tau} X^{\mu} (\tau,0) & = & \partial_{\tau} X^{\mu} (\tau,\pi) = 0,
\label{rev-Dirichlet-bc}
\end{eqnarray}
where we have parametrized the string length by $\sigma \in [0,\pi]$.
The Neumann boundary conditions are used to describe the free movement of the endpoint of the string. In contrast, the Dirichlet boundary conditions should be imposed when the string endpoint has a fixed coordinate in the space-time direction $\mu$. Then the conservation of energy and momentum in this direction imply that there is an object localized at the fixed value of the string endpoint transverse to the string. This object has several interesting physical properties as tension and charge under higher dimensional gauge fields (that generalize the massless vector field to fields described by higher dimensional differential forms) and it is called {\em D-brane}. The boundary conditions can be combined to Neumann-Neumann (NN), Dirichlet-Dirichlet (DD) and mixed Neumann-Dirichlet (ND) boundary conditions depending on the number of branes on which the string ends. 

The solutions to the equations of motion (\ref{rev-eq-X}) with all possible combinations of Neumann and Dirichlet boundary conditions are given by the following relations
\begin{eqnarray}
\mathrm{NN:} \hspace{0.5cm} X^{\mu}(\tau,\sigma) & = & x^{\mu} + 2{\alpha}'p^{\mu}\tau +
i\sqrt{2{\alpha}'}\sum_{n \neq
0}\frac{{\alpha}_{n}^{\mu}}{n}e^{-in\tau}\cos(n\sigma),
\label{rev-NN-bc}\\
\mathrm{DD:} \hspace{0.5cm} X^{\mu}(\tau,\sigma) & = & x^{\mu}_{0} + \frac{1}{\pi}(y^{\mu}_{0} - x^{\mu}_{0})\sigma +
\sqrt{2{\alpha}'}\sum_{n \neq
0}\frac{{\alpha}_{n}^{\mu}}{n}e^{-in\tau}\sin(n\sigma),
\label{rev-DD-bc}\\
\mathrm{ND:} \hspace{0.5cm} X^{\mu}(\tau,\sigma) & = & y^{\mu}_{0} + i\sqrt{2{\alpha}'}\sum_{r \in \,\mathbf{Z} +
\frac{1}{2}}\frac{{\alpha}_{r}^{\mu}}{r}e^{-ir\tau}\cos(r\sigma),
\label{rev-ND-bc}\\
\mathrm{DN:} \hspace{0.5cm} X^{\mu}(\tau,\sigma) & = & x^{\mu}_{0} + \sqrt{2{\alpha}'}\sum_{r \in \,\mathbf{Z} +
\frac{1}{2}}\frac{{\alpha}_{r}^{\mu}}{r}e^{-ir\tau}\sin(r\sigma).
\label{rev-DN-bc}
\end{eqnarray}
Here, we have denoted by $x^{\mu}_{0}$ and $y^{\mu}_{0}$ the values of the string endpoints, i. e. the ones fixed by the constraint \ref{rev-Dirichlet-bc}). The string fields are complex functions of the world-sheet coordinates. However, the embedding of the string in the space-time should be real. Therefore, one should impose the reality of the embedding components $X^{\mu}$ which amounts to imposing the {\em reality condition} on the Fourier coefficients, that is
\begin{equation}
\alpha^{\mu}_{-n} = \alpha^{\mu *}_{n}, \hspace{0.5cm} \beta^{\mu}_{-n} = \beta^{\mu *}_{n},
\label{rev-reality-cond}
\end{equation}
for all $n \in \mathbf{Z}$. Here, ${}^{*}$ denotes the complex conjugation. Similar reality conditions should be imposed on the open string fields.

The above solutions to the equations of motion are free massless fields $X^{\mu}$ in two dimensional spaces 
$\Sigma^2$. The  string degrees of freedom form a subset of $X^{\mu}$'s that is obtained after imposing the constraints
(\ref{rev-constraint-1}) and (\ref{rev-constraint-2}). Let us discuss in some detail the constraints of the 
closed strings. The constraints of the open strings with various boundary conditions can be analysed in the same way (see, e. g. \cite{Polchinski:1998rq}). Sometimes it is useful to introduce the {\em light-cone coordinates}
on the world-sheet by the linear transformations
\begin{equation}
\sigma^{\pm}   =  \tau \pm \sigma, \hspace{.5cm}
\partial_{\pm} = \frac{1}{2}\left( \partial_{\tau} \pm \partial_{\sigma}\right).
\label{rev-w-s-lc}
\end{equation}
Let us work out the string constraints in these coordinates. The equations of motion (\ref{rev-eq-X}) take the following form in the light-cone coordinates
\begin{equation}
\partial_{-}\partial_{+} X^{\mu} = 0.
\label{rev-eq-mot-lc}
\end{equation}
The closed string solution (\ref{rev-sol-cs}) can be decomposed into fields that describe the left- and right-moving modes
\begin{equation}
X^{\mu}(\sigma,\tau) = X_{L}^{\mu}(\sigma^{+}) + X_{R}^{\mu}(\sigma^{-}),
\label{rev-left-right-cs}
\end{equation}
where
\begin{eqnarray}
X_{L}^{\mu}(\sigma^{+}) &=& \frac{1}{2}x^{\mu} + \frac{1}{2}{\alpha}'p_{L}^{\mu}{\sigma}^{+} +
i\sqrt{\frac{\alpha'}{2}}\sum_{n \neq 0}{\frac{\beta_{n}^{\mu}}{n}e^{-in{\sigma}^{+}}},
\label{rev-L-cs}\\
X_{R}^{\mu}(\sigma^{-}) &=& \frac{1}{2}x^{\mu} + \frac{1}{2}{\alpha}'p_{R}^{\mu}{\sigma}^{-} +
i\sqrt{\frac{\alpha'}{2}}\sum_{n \neq 0}{\frac{\alpha_{n}^{\mu}}{n}e^{-in{\sigma}^{-}}}.
\label{rev-R-cs}
\end{eqnarray}
Here, the coefficients $p_{L}^{\mu} = p_{R}^{\mu}= p^{\mu}$. The constraints (\ref{rev-constraint-1}) and 
(\ref{rev-constraint-2}) take the following form
\begin{eqnarray}
\left( \partial_{+} X^{\mu}\right)^2 = \left( \partial_{-} X^{\mu}\right)^2 = 0,
\label{rev-constr-lc-ws}
\end{eqnarray}
and they represent the vanishing of the non-trivial components of the energy-momentum tensor in the light-cone coordinates on the world-sheet. When written in terms of the Fourier modes, the above equations take the following form
\begin{equation}
\sum_{n \in \mathbf{Z}} L^{\alpha}_{n} e^{-in\sigma^-} = \sum_{n \in \mathbf{Z}} L^{\beta}_{n} e^{-in\sigma^+} = 0,
\label{rev-Virasoro-constr-lc-ws}
\end{equation} 
where we have used the notation
\begin{equation}
L^{\alpha}_{n} = \frac{1}{2} \sum_{m \in \mathbf{Z}} \eta_{\mu \nu} \alpha^{\mu}_{n-m} \alpha^{\nu}_{m},
\hspace{0.5cm}
L^{\beta}_{n} = \frac{1}{2} \sum_{m \in \mathbf{Z}} \eta_{\mu \nu} \beta^{\mu}_{n-m} \beta^{\nu}_{m}.
\label{rev-Virasoro-gen-lc-ws}
\end{equation}
The Fourier coefficients of the zero modes are given by the following relation
\begin{equation}
\alpha^{\mu}_{0} = \beta^{\mu}_{0} = \sqrt{\frac{\alpha'}{2}} p^{\mu}.
\label{rev-Fourier-zero}
\end{equation}
Since the Fourier modes are independent of each other, the physical degrees of freedom in the Fourier representation should obey an infinite set of classical constraints
\begin{equation}
L^{\alpha}_{n} = L^{\beta}_{n} = 0,
\label{rev-classical-constr}
\end{equation}
for all $n \in \mathbf{Z}$. 
The objects $L^{\alpha}_{n}$ and $L^{\beta}_{n}$, are actually generators of an infinite number of symmetries that form the so called {\em classical Virasoro-Witt algebra} which completely determines the conformal field theory in two dimensions. The closed string displays two copies of this algebra. If the Virasoro algebra is imposed at the quantum level, then the quantum conformal field theory is solvable, too. Let us take a closer look at the generators $L^{\alpha}_{0}$ and $L^{\beta}_{0}$. From the defining equations
(\ref{rev-Fourier-zero}) we get
\begin{eqnarray}
L^{\alpha}_{0} & = & \frac{\alpha'}{2} p^{\mu} p_{\mu} +
\frac{1}{2} \sum_{m \in \mathbf{Z} \setminus \{0\} } \eta_{\mu \nu} \alpha^{\mu *}_{m} \alpha^{\nu}_{m} = 0,
\label{rev-L0-a}\\
L^{\beta}_{0} & = & \frac{\alpha'}{2} p^{\mu} p_{\mu} +
\frac{1}{2} \sum_{m \in \mathbf{Z} \setminus \{0\} } \eta_{\mu \nu} \beta^{\mu *}_{m} \beta^{\nu}_{m} = 0,
\label{rev-L0-b}
\end{eqnarray}
By comparing the above relations with the definition of the relativistic energy in $R^{1,d-1}$: $p^2 + M^2 = 0$ we see that the momentum of the center of mass of the closed string can be used to define the mass of string in terms of string modes
\begin{equation}
M^2 =  \frac{2}{\alpha'} \sum_{m \in \mathbf{Z}\setminus \{0\} } \eta_{\mu \nu} \alpha^{\mu *}_{m} \alpha^{\nu}_{m},
    =  \frac{2}{\alpha'} \sum_{m \in \mathbf{Z}\setminus \{0\} } \eta_{\mu \nu} \beta^{\mu *}_{m} \beta^{\nu}_{m}.
\label{rev-string-mass}
\end{equation}
The above relation is written in a form that reminds the quantum oscillator in the Fock space. Indeed, upon quantization and proper ordering of the Fourier operators, the relations (\ref{rev-string-mass}) can be interpreted as the definition of the mass operator of the quantum string. The theory is symmetric under the exchange of $\sigma^+ \leftrightarrow \sigma^-$ and the relabelling of the Fourier coefficients $\alpha \leftrightarrow \beta$. Consequently, the two sums that define the mass operator coincide at individual oscillator level, which is known as the {\em level matching condition}. We note once again that this condition is a consequence of the zero mode generators of the Virasoro algebra. The mass of the open string can be defined in a similar way. In this case, there is no level matching equation since the open string has just one Virasoro algebra.

\subsubsection{Canonical structure of string}

The classical string theory presented above suggest that one could attempt to quantize the $d$ scalar fields by canonical methods. The first step to apply these methods is to formulate the string in the phase space. Since there are constraints among string fields, the physical phase space is the subspace of the phase space determined by the equations  
(\ref{rev-constraint-1}) and (\ref{rev-constraint-2}). The string fields $X^{\mu}$'s and their canonically conjugate momenta
\begin{eqnarray}
P^{\mu} = \frac{\delta L}{\delta (\partial_{\tau}{X}_{\mu})} =
\frac{1}{2 \pi \alpha'}\partial_{\tau}{X}^{\mu},
\label{rev-conj-mom}
\end{eqnarray}
where $L$ is the string Lagrangian in the conformal gauge, satisfy the Poisson brackets at equal values of $\tau$ (equal-times)
\begin{eqnarray}
\{ X^{\mu}(\sigma),X^{\nu}(\sigma') \} &=&
\{ P^{\mu}(\sigma),P^{\nu}(\sigma') \} = 0
\label{rev-PB-1}\\
\{ X^{\mu}(\sigma),P^{\nu}(\sigma') \} &=& \delta(\sigma - \sigma'){\eta}^{\mu\nu}.
\label{rev-PB-2}
\end{eqnarray}
One can easily see that the above Poisson brackets imply the following relations among the classical Fourier coefficients 
\begin{eqnarray}
\{ \alpha_m^{\mu},\beta_n^{\nu} \} &=& 0 \label{rev-Fourier-PB-1},\\
\{ \alpha_m^{\mu},\alpha_n^{\nu} \} & = &
\{ \beta_m^{\mu},\beta_n^{\nu} \}  =  -im {\delta}_{m+n,0}{\eta}^{\mu\nu}.
\label{rev-Fourier-PB-2}
\end{eqnarray}
The coordinates of the center of mass satisfy the standard relations $\{ x^{\mu},p^{\nu} \} = {\eta}^{\mu\nu}$.
From the equations (\ref{rev-PB-2}) and (\ref{rev-Fourier-PB-2}), we can see that the String Theory has the same fundamental problem as the relativistic field theory, namely the temporal string field $X^{0}$ and its conjugate  momenta $P^{0}$ obey non-standard Poisson bracket relations with the negative sign induced by the $\eta^{00}$ component of the space-time metric. The same is true for the coordinates of the center of mass. 

The dynamics of the classical string in the phase space is determined by the Hamiltonian defined as
\begin{equation}
H =  \int d\sigma \left( P_{\mu} \partial_{\tau}{X}^{\mu} - L \right).
\label{rev-Hamiltonian-1}
\end{equation}
The explicit form of the Hamiltonian in terms of fields is
\begin{equation}
H = \frac{1}{4 \pi \alpha'} \int d\sigma\left[ (\partial_{\tau}X)^2 + (\partial_{\sigma}X)^2\right],
\label{rev-Hamiltonian-2}
\end{equation}
which vanishes on the constraint surface. Since the constraints are, in fact, equations of motion, one can also say that the Hamiltonian vanishes weakly, i. e. up to the equations of motion. A vanishing Hamiltonian on the surface of constraints, which is the physical subspace of the phase space, is a major problem for the theory. In general, that is solved by constructing a physical Hamiltonian in terms of physical degrees of freedom only (in this context the coordinates on the constraint surface) which obey a modified version of the Poisson brackets that takes into account the constraint structure. We are not going to analyse the string constraints here. More information about that can be found in \cite{Green:1987sp,Polchinski:1998rq}.

\subsection{String quantization}

The different methods to quantize the string can be told apart from each other by the way in which the constraints are implemented in the quantization procedure. The most consistent interpretation of the physical results by different observers is guaranteed by the covariant methods as in the case of the gauge field theories. However, in the study of the thermal properties of a system, several symmetries might be broken by the thermal effects which could be observer dependent. Therefore, in order to understand the dynamics of the thermal string, we will rely on the canonical quantization methods in which the constraints are firstly solved at classical level and then are imposed on the Fock space as relations among string operators.   

\subsubsection{Canonical covariant quantization}

As we have seen in the previous section, the closed string is described by a collection of $d$ massless scalar fields in two dimensions that form an $SO(1,d-1)$ massless vector in the space-time and are subjected to the constraints given by the equations (\ref{rev-constraint-1}) and (\ref{rev-constraint-2}). The scalar fields are free fields in the conformal gauge and, as the Poisson brackets (\ref{rev-Fourier-PB-1}) and (\ref{rev-Fourier-PB-2}) show, they have the canonical structure of a collection of oscillators in the spatial directions. In the time-like direction, the string modes are not physical as their norm is not positive definite. The canonical quantization method can be applied by promoting the string fields to operators and by replacing the Poisson brackets by commutators $\{ f(X,P) ,g(X,P) \} \rightarrow -i [\hat{f}(\hat{X},\hat{P}),\hat{g}(\hat{X},\hat{P} ) ]$. In order to simplify the notation, the hat will be omitted when the notation its presence is obvious. The Fock space representation is obtained by quantizing the Fourier modes of the fields. The equal time commutation relations take the standard form
\begin{eqnarray}
\left[X^{\mu}(\sigma),P^{\nu}(\sigma')\right] &=& i\delta(\sigma -
\sigma'){\eta}^{\mu\nu},
\label{rev-comm-rel-XP1}\\
\left[X^{\mu}(\sigma),X^{\nu}(\sigma')\right] &=&
\left[P^{\mu}(\sigma),P^{\nu}(\sigma')\right] = 0.
\label{rev-comm-relXP2}
\end{eqnarray}
By using the Fourier decomposition, one obtains the following commutation relations among the string oscillators 
\begin{eqnarray}
\left[{\alpha}_m^{\mu},\beta_n^{\nu}\right] &=&
0,
\label{rev-comm-rel-osc1}\\
\left[\alpha_m^{\mu}, \alpha_n^{\nu}\right] & = &
\left[\beta_m^{\mu},\beta_n^{\nu}\right]
= m{\delta}_{m+n}{\eta}^{\mu\nu}. 
\label{rev-comm-rel-osc2}
\end{eqnarray}
The coordinates of the center of mass satisfy the relation $\left[x^{\mu},p^{\nu}\right] = i{\eta}^{\mu\nu}$. By imposing the reality condition one obtains the following relations
\begin{equation}
\left(\alpha_{n}^{\mu} \right)^{\dagger} = \alpha_{-n}^{\mu},
\hspace{0.5cm}
\left(\beta_{n}^{\mu} \right)^{\dagger} =
\beta_{-m}^{\mu}.
\label{rev-quant-osc}
\end{equation}
The above equations show that the Fock space of the closed string is a direct product of an infinite number of harmonic oscillators corresponding to the modes $n$ and states of the center of mass. The oscillators are written in the string representation but they can be put into the canonical form by the rescaling 
\begin{equation}
\alpha^{\mu}_{n} \rightarrow a^{\mu}_{n} = \frac{\alpha^{\mu}_{n}}{\sqrt{n}},
\hspace{0.5cm}
\beta^{\mu}_{n} \rightarrow b^{\mu}_{n} = \frac{\beta^{\mu}_{n}}{\sqrt{n}},
\label{rev-can-osc}
\end{equation}
for all $n \neq 0$. One can see from the equation (\ref{rev-comm-rel-osc1}), that the Fock space is naturally factorized into Fock spaces corresponding to the left- and right-moving modes and of the momentum $p^{\mu}$. The vacuum state of the string is defined by the following relations
\begin{eqnarray}
\alpha^{\mu}_{n} \left| 0; p \right> & = &
\beta^{\mu}_{n}  \left| 0 ; p \right> = 0, \hspace{0.5cm} \forall n > 0,
\label{rev-vac-osc}\\
\hat{p}^{\mu} \left| 0;  p \right> & = & p^{\mu} \left| 0; p \right>.
\label{rev-vac-p}
\end{eqnarray}
An arbitrary state has contributions from both sectors with an arbitrary number of left- and right-string modes
\begin{equation}
\left| \phi ; p \right> = 
\prod_{r=1}^{\infty} 
\prod_{s=1}^{\infty}
\left( \alpha^{\mu_r} \right)^{m_r}_{n_r} 
\left( \beta^{\mu_s} \right)^{m_s}_{n_s} 
\left| 0 ; p \right>,
\label{rev-gen-state}
\end{equation}
where $r$ and $s$ indices indicate arbitrary string modes. Not all of the states of the form given by the relation (\ref{rev-gen-state}) are physical. The physical subspace of the Fock space is obtained by imposing the constraints (\ref{rev-constraint-1}) and (\ref{rev-constraint-2}) interpreted in terms of string operators. In the Fourier representation, the constraints take the form (\ref{rev-classical-constr}). One way to implement them is to impose the Virasoro-Witt generators (\ref{rev-classical-constr}) on the Fock space
\begin{equation}
L^{\alpha}_{n} \left|\phi ; p \right> = L^{\beta}_{n} \left|\phi ; p \right> = 0.
\label{rev-constr-Fock}
\end{equation}
for all $n \in \mathbf{Z}_{*}$, where
\begin{eqnarray}
L^{\alpha}_{n} & = & \frac{1}{2}\sum_{m = -\infty}^{\infty} \delta_{ij}\alpha^{i}_{n-m} {\alpha}^{j}_{m},
\label{rev-quant-La}\\
L^{\beta}_{n} & = & \frac{1}{2}\sum_{m = -\infty}^{\infty} \delta_{ij}\beta^{i}_{n-m} {\beta}^{j}_{m}.
\label{rev-quant-Lb}
\end{eqnarray}
However, the system (\ref{rev-constr-Fock}) is not self-consistent. Nevertheless, the weaker set of equations  
\begin{equation}
\left<\phi_{phys} ; p \right| L^{\alpha}_{n} \left| \phi_{phys} ; p \right> =
\left<\phi_{phys} ; p \right| L^{\beta}_{n} \left| \phi_{phys} ; p \right> = 0,
\label{rev-constr-weak}
\end{equation}
is properly defined and can be used to determine the physical states. The vanishing of the matrix elements of the Virasoro generators (\ref{rev-constr-weak}) is equivalent to taking $m \in \mathbf{Z}^{+}_{*}$. The zero generators of the zero modes should be normally ordered. The ordering procedure is ambiguous. If one defines the quantum zero mode operators by the following relations
\begin{eqnarray}
L^{\alpha}_{0} &=& \frac{1}{2} \alpha_{0}^{2} + 
\frac{1}{2}\sum_{n \neq 0} : \alpha_{-n} \cdot \alpha_{n}: = 
\frac{1}{2}{\alpha}_{0}^{2} + \sum_{n=1}^{\infty} \alpha_{-n} \cdot \alpha_{n},
\label{rev-ord1-L0-a}\\
L^{\beta}_{0} &=& \frac{1}{2} \alpha_{0}^{2} + 
\frac{1}{2}\sum_{n \neq 0} : \beta_{-n} \cdot \beta_{n}: = 
\frac{1}{2}{\alpha}_{0}^{2} + \sum_{n=1}^{\infty} \beta_{-n} \cdot \beta_{n},
\label{rev-ord1-L0-b}
\end{eqnarray}
a c-number should be introduced into the evaluation of the matrix elements on the physical space
because of the normal ordering ambiguity
\begin{eqnarray}
\left( L^{\alpha}_{0} - a_{\alpha} \right) \left| \phi_{phys} ; p \right> &=& 0,
\label{rev-ord-L0-a-q}\\
\left( L^{\beta}_{0} - a_{\beta} \right) \left| \phi_{phys} ; p \right> &=& 0.
\label{rev-ord-L0-b-q}
\end{eqnarray}
The generators satisfy the quantum Virasoro algebra 
\begin{equation}
\left[L_{m},L_{n} \right] = (m-n)L_{m+n} + \frac{c}{12}m(m^2-1) {\delta}_{m+n,0},
\label{rev-Virasoro-alg}
\end{equation}
where $L$ stands for the generators from either left- or right-moving sector, respectively, and $c$ is the {\em central charge}. In order to remove the negative norm states from the physical subspace, the central charge must have a unique integer value $c = 26$ \cite{Green:1987sp}. Since $c$ counts the number of scalars, it follows that $d = c = 26$. Thus, the condition to have a ghost free theory fixes uniquely the space-time dimension. The set of generators $L_{-1}, L_0, L_{1}$ form the sub-algebra $sl(2,\mathbf{Z})$ of the Virasoro algebra. The closed string has two copies of the Virasoro algebra corresponding to the left- and right-sector, respectively, and two copies of the $sl(2,\mathbf{Z})$ algebra.

\subsubsection{Light-cone quantization}

In order to solve the constraints given by the equations (\ref{rev-constraint-1}) and (\ref{rev-constraint-2}) one can choose the {\em light-cone gauge} in space-time defined by the following linear transformations
\begin{equation}
X^{\mu} \longrightarrow X^+, X^-, X^i,
\label{rev-light-cone-st-tr}
\end{equation} 
where $i = 1,2,\ldots , 24$ and 
\begin{equation}
X^{\pm}=\sqrt{\frac{1}{2}}\left( X^{0} \pm X^{d-1} \right).
\label{rev-light-cone-st-01}
\end{equation}
To the light-cone directions one can associate momenta $P^{\pm}$. The light-cone gauge is well defined when $P^+$ is non-zero ($P^-$ can be expressed as a function on $P^+$ and $P^i$'s) and it is not a covariant gauge. However, by imposing the Lorentz invariance of the results obtained in the light-cone gauge, one can obtain useful information about the system. Also, in this gauge the spectrum is ghost free.

The space-time metric in the light-cone gauge has the following form
\begin{equation}
ds^2 = -2dX^+ dX^- + \sum_{i,j=1}^{24} \delta_{ij} dX^i dX^j.
\label{rev-metric-lcg}
\end{equation}
The string field $X^+$ can be chosen to have the following form
\begin{equation}
X^+ = x^+ + 2 \alpha' p^+ \tau.
\label{rev-X-plus}
\end{equation}
Since $X^+$ is fixed, one has to solve the equations of motion of other fields and to determine their Fourier expansion. The computations are straightforward and can be found in the references, e. g. \cite{Green:1987sp}. For example, by quantizing the open string with NN boundary conditions, one obtains the following relation among the string oscillators
\begin{equation}
\alpha_{m}^{-} = 
\sqrt{\frac{1}{2\alpha'}} \, p^{+}
\left\{\frac{1}{2}\sum_{i=1}^{d-2}\sum_{ n=-\infty}^{\infty}\delta_{ij}: \alpha_{m-n}^{i} \alpha_{n}^{j}:
-a \delta_{n,0}\right\}.
\label{rev-op-st-alpha-lcg}
\end{equation}
From it, one can derive the zero mode Virasoro generator 
\begin{equation}
L_{0} = -2{\alpha}' \left( p^{+}p^{-} + \frac{1}{2\alpha'} N + \frac{1}{2} \sum_{i,j=1}^{24} \delta_{ij} p^{i}p^{j} \right).
\label{rev-Vir-lcg}
\end{equation}
The string excitation are classified by their mass. The mass operator is defined by the quantum version of the equation
(\ref{rev-string-mass}) truncated to the open string and it is given by the following relation
\begin{equation}
M^2 = \frac{4}{\alpha'} \left( N - a \right),
\label{rev-mass-op-open}
\end{equation}
where $a$ is the c-number from the normal ordering and the number operator has the following form 
\begin{equation}
N = \sum_{i=1}^{d -2 = 24} \sum_{m} \delta_{ij} \alpha^{i}_{-m} \alpha^{j}_{m}.
\label{rev-number-op}
\end{equation}
Similar relations can be obtained for the rest of the boundary conditions of the open string as well as for the closed string.
The quantum mass operator of the closed string has the following form
\begin{equation}
M^2 = \frac{2}{\alpha'}\left( N^{\alpha} + N^{\beta} - a_{\alpha} - a_{\beta} \right),
\label{rev-mass-op-cs}
\end{equation}
where $N^{\alpha}$  and $N^{\beta}$ stand for the number operator in the corresponding sectors
\begin{equation}
N^{\alpha} = \sum_{i=1}^{d -2 = 24} \sum_{m} \delta_{ij} \alpha^{i}_{-m} \alpha^{j}_{m},
\hspace{0.5cm}
N^{\beta} = \sum_{i=1}^{d-2 = 24} \sum_{m} \delta_{ij} \beta^{i}_{-m} \beta^{j}_{m}.
\end{equation}
The physical states satisfy the level matching condition
\begin{equation}
\left<\phi_{phys} ; p \right| L^{\alpha}_{0} - L^{\beta}_{0} - a_{\alpha} - a_{\beta}
 \left| \phi_{phys} ; p \right> = 0.
\label{rev-quant-level-match}
\end{equation}
The open string has only one copy of the Virasoro algebra, therefore no level matching condition is necessary. The light-cone gauge can be used to establish the space-time dimension and the value of the normal ordering c-numbers. Also, one can determine easily the physical states in this gauge.

\subsubsection{Physical spectrum}

By using either the canonical quantization or the light-cone quantization one can calculate the physical spectrum of the closed bosonic string. Let us recall the states of the closed string. A general state can be written as
\begin{equation}
|\phi \rangle_{cs} = |\phi_{\alpha},\phi_{\beta}\rangle  = 
|\phi_{\alpha}\rangle \otimes |\phi_{\beta} \rangle 
\label{rev-cs-gen-state}
\end{equation}
The entangled states on the boundary of the world-sheet in general do not belong to the physical Fock space since they do not have finite norm. The number of left- and right-moving operators in each sector should match by the level matching condition which can be written as
\begin{equation}
\left( N^{\alpha} - N^{\beta} \right) |\phi \rangle_{cs} = 0.
\label{rev-lev-match-state}
\end{equation}
The relation (\ref{rev-lev-match-state}) shows that the first excited states are of the form
\begin{equation}
\zeta_{\mu\nu} \alpha_{-1}^{\mu} \beta_{-1}^{\nu}|0;p \rangle,
\hspace{0.25cm}
p^{\mu }{\zeta}_{\mu\nu} = p^{\nu }{\zeta}_{\mu\nu} = 0,
\hspace{0.25cm}
M^2 \alpha_{-1}^{\mu} \beta_{-1}^{\nu}|0;p \rangle = 0,
\label{rev-massless-cs}
\end{equation}
where the second relation above is the result of the following equations  
\begin{eqnarray}
L^{\alpha}_{1} \zeta_{\mu \nu} \alpha_{-1}^{\mu} \beta_{-1}^{\nu} \left|0;p\right> & \sim &
\zeta_{\mu \nu}p^{\mu}\left|0;p\right>,
\label{rev-act-L1}\\
L^{\beta}_{1} \zeta_{\mu \nu} \alpha_{-1}^{\mu} \beta_{-1}^{\nu} \left|0;p\right> & \sim &
\zeta_{\mu \nu}p^{\nu}\left|0;p\right>.
\label{rev-act-L2}\\
\end{eqnarray}
The equations (\ref{rev-act-L1}) and (\ref{rev-act-L2}) represent gauge transformations in the Fock space. The two equations leave arbitrary only the transverse components of the general tensor $\zeta_{\mu \nu}$. Thus, the massless string states are parametrized by $\zeta_{(ij)}, \zeta_{[ij]}, \mbox{Tr}{\zeta}$ and are classified by the irreducible representations of $SO(24)$. By determining the transformation of these excitations under the Lorentz group one can see that they correspond to the graviton $h_{ij}$, the antisymmetric (Kalb-Ramond) field $B_{ij}$ and to the dilaton $\phi$.
In particular, the graviton transforms under 
\begin{equation}
h_{ij} \longrightarrow h_{ij} + \partial_{(i} \zeta_{j)},
\label{rev-grav-transf}
\end{equation}
where $\zeta_i$ is an arbitrary infinitesimal vector from $SO(24)$. The natural appearance of the graviton in the string spectrum, from quantum consistency and symmetry considerations, is one of the strongest motivation for pursuing the research in the field of the String Theory. 

The physical spectrum of the open string contains the ground state $|0;p\rangle$ which represents the lowest mass string excitation and has negative mass $M^2=-\frac{a}{\alpha'}$. Thus, the ground state shows that there is a tachyonic field $T$
in the theory. The standard interpretation of this fact is that the theory has been quantized around an unstable vacuum. The next excitation in the open string spectrum  is of the form $\zeta_{i} {\alpha}_{-1}^{i} |0;p\rangle$ of mass $M^2 = (1-a) / \alpha'$ and $\zeta_i p^i = 0$ from the gauge transformation generated by $L_1$. This is an $SO(24)$ vector and it has transversal degrees of freedom only for $M^2 = 0$, which fixes the value of the normalization c-number to one. Thus massless vector fields show up in the string theory, too. The presence of all massless fields that mediate the fundamental interactions makes the String Theory a candidate for a theory that unifies these interactions.

The massive states start at values proportional to ${\alpha'}^{-1} \sim 10^{36} GeV $. The study of these states is much less advanced and their physical properties, such as their renormalized mass, are still largely unknown.

\subsubsection{D-brane states}

In this subsection we briefly review the construction of the D-brane states as boundary states in the perturbative string theory. There are many good references in the literature. The presentation follows 
\cite{Polchinski:1998rq,Johnson:2005mqa,Szabo:2004uy,DiVecchia:1999rh} and mainly the fifth chapter from \cite{Vancea:2001hj}.

At the tree level, the free closed string is generated from vacuum, propagates a while and then is annihilated to
vacuum. The corresponding diagram is thus a cylinder which is the string world-sheet. The states that describe the annihilation and creation of the string are located at the end circles of the cylinder at the initial and final instants of time, respectively. These states are called {\em boundary states}. The same cylinder can be interpreted as being the one-loop diagram of a free open string. The cylinder in the open string sector is different from the one in the closed string sector in that the space-like coordinates on the two world-sheets get interchanged. This coordinate transformation is implemented to the string fields and generates a the open-closed string duality.  In the open string description, the end points of the string can move on some hypersurfaces located in space-time at loci that correspond to the location of the two end circles. These hypersurfaces are transversal to the string. The energy and momentum is conserved only if there are extended transversal physical objects that are extended along the hypersurfaces and exchange momentum and energy with the string. These objects are the {\em D-branes} and they generalize the strings to higher dimensional objects. A $D$-brane with $p$ space-like coordinates is called a $Dp$-brane. It describes a $p+1$-dimensional hypersurface $\Sigma^{p+1} \subset\mathbf{R}^{1,9}$ called the {\em world-volume} of the $Dp$-brane.

The open string boundary conditions that define a $Dp$-brane are given by the following relations
\bea
\6_{\s}X^{a}|_{\s=0} & = & 0,~~~a = 0,1,\ldots ,p \nonumber\\
X^{i}|_{\s =0} & = & y^{i}, ~~~i = p+1, \ldots ,24. 
\label{bcopenstr}
\eea
To pass to the closed string boundary condition, one has to interpret the 
cylinder as the tree-level diagram in the closed string sector. The relations (\ref{bcopenstr}) take the following form
\bea
\6_{\t}X^{a}|_{\t=0}&=&0,~~~a=0,1,\ldots ,p\nonumber\\
X^{i}|_{\t =0}&=&y^{i},~~~i=p+1,\ldots\ 25.
\label{bcclosedstr}
\eea
If we want to describe the $Dp$-branes as boundary states, we must 
implement the boundary conditions (\ref{bcclosedstr}) in the Fock space of the 
perturbative closed string
\bea
\6_{\t}X^{a}|_{\t=0}|B>&=&0,~~~a=0,1,\ldots ,p\nonumber\\
(X^{i}|_{\t =0}- y^i)|B>&=&0,~~~i=p+1.\ldots\ 25.
\label{bchilbert}
\eea
The equations (\ref{bchilbert}) define the boundary state $|B>$. To find their
solution, we expand the string operators in terms of oscillation modes using
the solution of the equations of motion given in the previous section
\be
X^{\m}(\t ,\s )=x^{\m} + 2 \alpha' p^{\m} \t + i \sqrt {\fr{\alpha'}{2}} 
\sum^{\infty}_{n \neq 0} \left [\alpha^{\m}_{n}e^{-2in(\t - \s)}+
\beta^{\m}_{n}e^{-2in(\t + \s)} \right ].
\label{oscexpansion}
\ee
The string excitations are obtained by acting with products of creation operators on the vacuum state
\be
|0>=|0>_{\alpha}|0>_{\beta}|p>.
\ee
By substituting the expression (\ref{oscexpansion}) into the equation
(\ref{bchilbert}), one obtains the following boundary equations
\bea
(\alpha^{a}_{n} + \beta^{a}_{-n})|B>&=&0,    \nonumber \\
(\alpha^{i}_{n} - \beta^{i}_{-n})|B>&=&0,    \nonumber \\
\hat{p}^{a}|B>&=&0,   \nonumber \\
(\hat{x}^{i} - y^{i})|B>&=&0.
\label{bcosc}
\eea
Note that the equations (\ref{bcosc}) are not the only conditions that
should be imposed on the Hilbert space. Actually, one has to solve for 
{\em physical boundary states}, and therefore the negative norm states should
be excluded from the solutions of the system (\ref{bcosc}). In the light-cone gauge 
these equations are satisfied by the physical degrees of freedom. It follows that
the boundary states have the following general form
\be
|B_{X}>=N_{p}\d^{25-p}(\hat{x}^{i}-y^{i}) \left ( \prod^{\infty}_{n=1} 
e^{- \fr{1}{n} \alpha_{-n} \cdot S \cdot \beta_{-n}} \right )|0>_{\alpha} |0>_{\beta}|p=0>,
\label{Dpboundstate}
\ee
where $N_p$ is a normalization constant that should be determined and the matrix
$S$ has the following form
\be
S = (\eta^{ab},-\delta^{ij}).
\label{smatrix}
\ee 
In the equation (\ref{Dpboundstate}) the light-cone gauge implies the 
summation over the 24 transverse directions on which the metric is 
Euclidean. In order to have a complete knowledge of the $Dp$-brane state, one has to 
calculate the normalization constant $N_p$. This can be done by comparing the
interaction amplitudes computed in the closed and open string channels, respectively, (see, e. g. \cite{pol,DiVecchia:1999rh,Vancea:2001hj}). 
After lengthy computations, the result has the form
\be
N_{p}=\fr{T_{P}}{2},
\label{normconst}
\ee
where $T_p$ is the brane tension.

\section{Thermo Field Dynamics}

In this section, we are going to review the main results of the Thermo Field Dynamics (TFD) formalism for studying the 
thermal properties of quantum fields. These properties are  difficult to be studied from the first principles, i. e. from the microscopic interaction of fields with the environment or with each other \cite{hu}.

The application of the String Theory to cosmological problems such as the study of the primordial universe and the inflation requires understanding the behavior of strings in contact with the environment. Since the strings are supposed to be fundamental objects, this contact should be a very complex interaction among systems composed by large numbers of strings. In general, the description of this type of interaction in terms of individual strings is impossible. Therefore, it is necessary to develop statistical methods for string systems. The most rigorous way to do that is by developing a String Field Theory and then study it in interaction with the environment. There have been several attempts to do that \cite{Leblanc:1987zj,Abdalla:2005qs,Arafeva:2015} but the results are far from being satisfactory. The main reason for that is our insufficient understanding of the String Field Theory. However, there is a class of systems in which the contact between the strings and the environment can be described as a problem of equilibrium with a thermal reservoir. In these cases, it should be possible to formulate the interaction in terms of a field theory at finite temperature. 

In the case of strings, one should make the distinction between the field theories on the two dimensional world-sheets that can be associated to an individual string or several interacting strings and the String Field Theory which describes a large number of strings. The thermal effects should be understood in both theories. In this paper, we are going to study the thermal effects on the world-sheet.

It is possible to introduce the temperature in either operatorial or path integral formalisms. In this paper, we are going to present only the first method called the Thermo Field Dynamics (TFD) following closely \cite{hu}. While it is completely equivalent with the path integral formulation at the thermodynamical equilibrium , it has the advantage of preserving the non-compact time direction and maintaining explicit the structure of string excitations which is useful when one likes to identify the pattern of the symmetry breaking by temperature.

\subsection{Thermal vacuum}

The Thermo Field Dynamics proposed in \cite{ut} (see also \cite{hu,Khanna:2009zz}) has as a starting point the interpretation of the statistical average of an hermitian operator $A$ as a vacuum expectation value in a paricular state called the {\em thermal vacuum} $|0(\b) \ra $ 
\begin{equation}
\la 0(\b)|A|0(\b) \ra  = \frac{1}{Z(\b)} \mbox{Tr} \left[ e^{-\b H' } A \right]
\label{rev-0},
\end{equation}
where 
\begin{equation}
H' = H-\mu N, \hspace{0.5cm} Z(\b)=\mbox{Tr}\left[e^{-\b H' } \right],
\label{rev-not}
\end{equation}
and $\b=\frac {1}{{k}_{B}T}$. Here, $H$ is the  Hamiltonian, $\mu$ is the chemical potential, $N$ 
is the number of particles and ${k}_B$ is the Boltzmann constant. The key concept of this construction is the thermal vacuum. Consider, for simplicity, that the Hamiltonian has a discrete spectrum $H'|n \ra =\o_{n}|n\ra$ with orthogonal normalized states $\la n|m \ra=\d_{nm}$. Then the right hand side of the equation (\ref{rev-0}) takes the form
\begin{equation}
\la 0(\b)|A|0(\b) \ra=Z^{-1}(\b) \sum_{n}\la n|A|n \ra e^{-\b {\o}_{n}}. 
\label{rev-1111}
\end{equation}
On the other hand, since the eigenstates form a basis of the Hilbert space ${\mathcal{H}}$, one can make the expansion
\begin{equation}
|0(\b) \ra=\sum _{n}|n \ra \la n |0(\b) \ra=\sum _{n}f_{n}(\b)|n \ra .    
\label{rev-2a}
\end{equation}
Now it is easy to see that the coefficients $f_{n}(\b)$ must satisfy the following equation
\begin{equation}
f^{\ast}_{n}(\b)f_{m}(\b) = Z^{-1}(\b)e^{-\b \o_{n}} \d_{nm}, 
\label{rev-3a}
\end{equation}
which shows that $f_{n}(\b)$ are vectors rather than complex numbers. Consequently, the thermal vacuum $|0(\b) \ra$ belongs to an extended Hilbert space that contains a copy of every degree of freedom of the original system. The above construction shows that the doubling of the Hilbert space is a natural and necessary step when one studies a thermal system. The copy of the original Hilbert space 
$\mathcal{H}$ is denoted by $\tilde{\mathcal{H}}$. Also, we denote by tilde any quantity that reefers to the copy. The tilde system is interpreted as (that part of) the thermal reservoir that interacts with the original system and thermalizes it. The two systems are independent of each other apart for the interaction that leads to the thermalization of the original system. Thus, we can write
\begin{equation}
\widetilde {\cal{H}}| \widetilde n \rangle =\o_{n}| \widetilde n \ra, \hspace{0,5cm} 
\langle  \widetilde n| \widetilde m \ra =\d_{mn},
\label{rev-tilde-osc}
\end{equation}
where $\o_{n}$'s are the same frequencies as of the original oscillators.
The {\em total system} is formed by the original system and its copy from the thermal reservoir. Since the two are independent from each other, a total state is the direct product of one state from each subsystems of the total system. For example, the basis vectors have the form $|n, \widetilde m \ra=|n\rangle \otimes | \widetilde m \ra$. In this basis, the Hermitian operators $A$ and $ \widetilde A$ 
associated to a certain physical quantity have the following matrices
\begin{equation}
 \langle  \widetilde m,n|A|n', \widetilde m' \ra = \langle n|A|n' \ra \d_{mm{'}}, \ \ \langle  \widetilde m,n| \widetilde A|n', \widetilde m' \ra = \langle  \widetilde m| \widetilde A| \widetilde m' \ra \d_{nn{'}}.
\label{rev-matrices-AtildeA}
\end{equation}
One can easily check that the coefficients $f_{n}(\b)$ from the equation ({\ref{rev-2a}}) are given in terms of vectors from $\tilde{\mathcal{H}}$ by the following relation
\begin{equation}
f_{n}(\b)=e^{-\b \o_{n}/2}Z^{-1/2}(\b)| \widetilde n \ra.   
\label{rev-def}
\end{equation}
This fixes the form of the thermal vacuum to
\begin{equation}
|0(\b) \ra = \sum_{n}e^{-\b \o_{n}/2}Z^{-1/2}(\b) |\widetilde n \ra \otimes | n\rangle = \sum_{n}e^{-\b \o_{n}/2}Z^{-1/2}(\b)|n,\widetilde n \ra. \label{rev-vab}
\end{equation}
The normalization of the thermal vacuum to unity $\langle 0(\b)|0(\b) \ra = 1$ gives the standard expression for the partition function
\begin{equation}
Z(\b)=\sum_{n}e^{-\b \o_{n}} \langle n|n\rangle=\mbox{Tr}\;[e^{-\b {\cal{H}}}].
\label{rev-part-fct-stand}
\end{equation}
Then the vacuum expectation value in the thermal vacuum of an operator $A$ is given by the following equation
\begin{equation}
 \langle 0(\b)|A|0(\b) \ra = Z^{-1}(\b) \sum_{n}e^{-\b \o_{n}} \langle n|A|n \ra . 
\label{ref-vev-A}
\end{equation}
By duplicating the system and defining the thermal vacuum which formalizes the process of thermalization, one obtains a consistent quantum field theory at finite temperature in which all the standard techniques of the quantum field theory can be applied.

\subsection{Thermal harmonic oscillator}

The simplest system to which the above formalism can be applied is the harmonic oscillator. This example is instructive as it shows how to construct the thermal Fock space of a free boson. Let 
$a$ and $a^{\a}$ be the annihilation and creation oscillator operators
\begin{equation}
{\bo{H}}=\o a^{\a}a, \hspace{0.5cm}
[a,a^{\a}]=aa^{\a}-a^{\a}a=1, \hspace{0.5cm} 
[a,a]=[a^{\a},a^{\a}]=0.  \label{rev-com1}
\end{equation}
The eigenvalue problem is defined by the standard equation
\begin{equation}
{\bo{H}}|n \ra =\o_{n}|n\rangle \hspace{0.5cm} n=0,1,2, \dots, \infty,    
\end{equation} 
where $\o_n = n \o$ for all $n > 0$. To the harmonic oscillator we associate its tilde copy from the thermal reservoir defined by the equations
\begin{equation}
\widetilde{\bo {H}}=\o  \widetilde a^{\a}  \widetilde a,
\label{rev-tilde-osc-bos}
\end{equation}
with the commutation algebra
\begin{equation}
[ \widetilde a,  \widetilde a^{\a}]=1, 
\hspace{0.5cm} 
[ \widetilde a, \widetilde a]=[ \widetilde a^{\a}, \widetilde a^{\a}]=0. 
\label{rev-tilde-osc-alg}
\end{equation}
The two oscillators are independent of each other
\begin{equation}
[a, \widetilde a]=[a^{\a}, \widetilde a^{\a}]=[a, \widetilde a^{\a}]=[a^{\a}, \widetilde a]=0.
\label{rev-osc-tilde-osc-ind}
\end{equation}
The states of the total system are given by linear combinations of the following vectors
\begin{equation}
|0\ra \! \ra, \ \ \ a^{\a}|0 \ra \! \ra, \ \ 
\widetilde a^{\a}|0 \ra \! \ra, \ \    
a^{\a} \widetilde a^{\a}|0 \ra \! \ra,\ \   
\frac{1}{n!}(a^{\a})^{n}( \widetilde a^{\a})^{n}|0 \ra \! \ra, \dots  
\label{rev-efd}
\end{equation}
where the vacuum of the total system at zero temperature is given by the following relation
\begin{equation}
|0, \widetilde{0}\ra =|0 \ra \otimes \w {|0 \ra} = |0\ra \! \ra.
\label{rev-not-vac}
\end{equation}
The thermal vacuum can be constructed from the above states of the doubled or total system by applying the relations from (\ref{rev-vab}) and it takes the following form
\begin{equation}
|0(\b)\ra = Z^{-1/2}(\b)
\sum_{n}\frac{e^{-\b \o_{n}/2} }{n!}(a^{\a})^{n}( \widetilde a^{\a})^{n}|0 \ra \! \ra. 
\label{rev-vtb}
\end{equation}
The normalization of $|0(\b)\ra$ to unity gives the known expression for the partition function
\begin{equation}
Z(\b)= \frac{1}{1-e^{-\b \o}} .
\label{rev-part-bos}
\end{equation}
This fixes the phase of the thermal vacuum given by the equation (\ref{rev-vtb}) and which now takes the following form 
\begin{equation}
|0(\b)\ra = \sqrt{1-e^{-\b \o}} \exp \le(e^{-\frac{\b}{2}\o} a^{\a} \widetilde a^{\a} \ri)|0 \ra \! \ra . 
\label{rev-vac-bos-osc}
\end{equation}
The above equation shows that the thermal vacuum is an entangled state of states of the system and the tilde system at zero temperature. It is interesting to explore further the relationship between the doubled system and the thermal one.

\subsubsection{Bogoliubov transformations}

The equation (\ref{rev-vac-bos-osc}) can be interpreted as a transformation from the total vacuum at zero temperature to the thermal vacuum generated by a Bogoliubov operator $G(\th)$ of the form
\begin{equation}
G(\th)=G(\th)^{\a}=-i\th(\b)( \widetilde{a}a- {a}^{\a}\widetilde{a}^{\a}),  
\label{rev-gtb}
\end{equation}
where $\theta(\b)$ is a temperature dependent real parameter defined by any of the following relations 
\begin{eqnarray}
u(\b) & = & (1-e^{-\b \o})^{-\frac{1}{2}}=\sqrt{1+f_{B}(\o)}=\cosh\th(\b),  \label{rev-u11}
\\
v(\b) & = &(e^{\b \o}-1)^{-\frac{1}{2}}=\sqrt{f_{B}(\o)}=\sinh\th(\b).  \label{rev-v11}
\end{eqnarray}
Here, $f_{B}$ is the Bose-Einstein distribution. The operator $G(\th)$ is Hermitian and satisfies the following algebra 
\begin{equation}
[G,a] = -i\th (\b) \widetilde{a}^{\a}, \hspace{0.5cm} [G, \widetilde{a}]=-i\th (\b){a}^{\a}, 
\label{rev-alg-Bog-1}
\end{equation}
\begin{equation}
[G,a^{\a}]  = -i\th (\b) \widetilde{a}, \hspace{0.5cm} [G, \widetilde{a}^{\a}]=-i\th (\b)a.
\label{rev-alg-Bog-2}
\end{equation}
It follows that the operator $G(\th)$ generates the unitary Bogoliubov transformation 
$U(\th)=e^{-iG(\th)}$ which maps the total Hilbert space at zero temperature in to the Hilbert space at finite temperature. By this transformation,
the total vacuum is mapped to the thermal vacuum as
\begin{equation}
|0(\b)\ra =e^{-iG(\th)}|0\ra \! \ra .  \label{rev-tbv}
\end{equation}
The Bogoliubov transformation has the following action on the creation and annihilation operators 
\begin{eqnarray}
a(\b)&=&e^{-iG}ae^{iG} = u(\b)a-\v(\b) \widetilde{a}^{\a} \label{rev-aa} \\ 
\widetilde{a}(\b)&=&e^{-iG} \widetilde{a}e^{iG} = u(\b) \widetilde{a}-\v(\b){a}^{\a}, \label{rev-at} \\
a^{\a}(\b) & = & e^{-iG}ae^{iG}=u(\b)a^{\a}-\v(\b) \widetilde{a} , \label{rev-atd} \\
\widetilde{a}^{\a}(\b) & = & e^{-iG} \widetilde{a}^{\a}e^{iG}=u(\b) \widetilde{a}^{\a}-\v(\b){a}. \label{rev-atd1}
\end{eqnarray}
These equations establish the relationship between the Hermitian conjugation and the tilde conjugation viewed as internal operations in the algebra of the operators of the total system.

\subsubsection{Thermal Fock space}

By using the above relations, one can prove that the thermal vacuum satisfies the following equations
\begin{eqnarray}
a(\b)|0(\b)\ra =0, 
\label{rev-av} \\
\widetilde{a}(\b)|0(\b)\ra = 0.
\label{rev-avt}
\end{eqnarray}
This shows that the thermal vacuum contains no thermal excitations which justifies the name of the state
$|0(\b)\ra$. The annihilation operators at zero temperature act on the thermal vacuum as follows
\begin{eqnarray}
a|0(\b)\ra & = & e^{-\b{\hat{H}} /2}{\w{a}}^{\a}|0(\b)\ra ,\label{rev-vou}\\
\w{a}|0(\b)\ra & = & e^{-\b{\hat{H}} /2}{a}^{\a}|0(\b)\ra.\label{rev-vou1}
\end{eqnarray} 
The Fock space of the thermal oscillator can be constructed by acting with the operators $a^{\a}(\b)$and $\widetilde{a}^{\a}(\b)$ on the thermal vacuum $|0(\b)\ra$. However, these states are not eigenstates of any of the two Hamiltonian operators of either the harmonic oscillator or the tilde oscillator. Nevertheless, it is easy to verify that the new states are eigenstates of the following total Hamiltonian
\begin{equation}
\hat{H}= H- \widetilde{H}=\o \le(a^{\a}a- \widetilde{a}^{\a} \widetilde{a} \ri)=\o \le( a^{\a}(\b)a(\b)-\widetilde{a}^{\a}(\b) \widetilde{a}(\b) \ri ).  
\label{rev-ht}
\end{equation}
The total Hamiltonian commutes with the generator of the Bogoliubov transformation $G(\th)$. Therefore, the thermal structure is preserved under the evolution generated by the total Hamiltonian. The physical quantities at finite temperature are given by the vacuum expectation value of the original observables in the thermal vacuum, as stated above. For example, the number of quanta at finite temperature is computed by the vacuum expectation value of the number operator
\begin{equation}
 \langle 0(\b)|{a}^{\a}a|0(\b)\ra = \frac{1}{e^{\b \o}-1}, \label{rev-n}
\end{equation}
which is just the Fermi-Boson distribution as expected.

\subsubsection{Thermal doublet}

 It is useful to tensor the operators with the $\mathbf{C}^2$. The resulting formulation is called the 
{\em thermal doublet} formulation of the TFD. The basic doublet of the string oscillators at zero temperature is the following operator
\begin{equation} 
A=\le(\begin{array}{c} a \\ \w{a}^{\a} \end{array}  \ri ).
\end{equation}
By applying the unitary Bogoliubov transformation to it, one obtains the corresponding thermal doublet
\begin{eqnarray}
A(\b) = \le(\begin{array}{c} a(\b) \\ \w{a}^{\a}(\b) \end{array}  \ri )=U(\th)AU^{\a}(\th)
= \fourmat{\cosh {\th(\b)}}{-\sinh {\th(\b)}}{-\sinh {\th(\b)}}{\cosh {\th(\b)}} 
\le(\begin{array}{c} a \\ \w{a}^{\a} \end{array}  \ri ),
\end{eqnarray}
where the equations (\ref{rev-aa}) and (\ref{rev-atd1}) have been used.

\subsection{Thermal bosonic free field}

The TFD method can be extended to collection of oscillators, in particular to free fields at finite temperature. The original formulation was made for non-relativistic fields \cite{hu}, but the method can be applied to relativistic fields as well \cite{Khanna:2009zz}. For simplicity, we are going to review the non-relativistic bosonic field and follow \cite{hu}.

\subsubsection{Total system}

The original Lagrangian of the bosonic field $\psi({\ve{x}},t)$ in $R^{1+3}$ is given by the following equation
\begin{equation}
{\cal{L}}(\ve{x},t)=i\psi^{\ast}\dot{\psi}-\frac{1}{2m}
{\boldmath{\nabla}}{\psi}^{\ast}{\boldmath{\nabla}}{\psi}.
\label{rev-lag-sch}
\end{equation}
The field and its complex conjugate are independent fields that obey the Schr\"{o}dinger equation. To this field one associates the field $\widetilde{\psi}$ corresponding to the degrees of freedom of the thermal reservoir and with the following Lagrangian
\begin{equation}
{ \widetilde{\cal{L}}}(\ve{x},t)=-i {\widetilde{\psi}}^{\ast}{ \dot{\widetilde{\psi}}}-\frac{1}{2m}{\boldmath{\nabla}} {\widetilde{{\psi}}}^{\ast}{\boldmath{\nabla}} \widetilde{{\psi}}.
\label{rev-lag-sch-tilde}
\end{equation}
The two systems can be quantized independently of each other by applying the standard canonical quantization method. The non-vanishing equal-time commutation relations are
\begin{eqnarray}
\left[\hat{\psi}({\ve {x}},t),\hat{\psi}^{\a}({\ve {x}}{'},t) \right] & = & \d^{3}({\ve {x}} - {\ve {x}}{'}),
\label{rev-comm-psi}
\\
\left[ \widetilde{\psi}({\ve {x}},t), \widetilde{\p}(\ve{x}',t) \right] & = & \d^{3}({\ve {x}} - \ve{x}').
\label{rev-comm-tilde-psi}
\end{eqnarray}
Since the two system are independent 
\begin{equation}
\left[ {\psi}({\ve {x}},t), \widetilde{\psi}(\ve {x}',t) \right] 
=
\left[  \widetilde{\psi}({\ve {x}},t),{\psi}^{\a}(\ve{x}',t) \right]=0.
\label{rev-comm-psi-tilde-psi}
\end{equation}
The Lagrangian of the total system ${\hat{\cal{L}}}(\ve{x},t)$ is defined by the following equation
\begin{equation}
{\hat{\cal{L}}}(\ve{x},t)={\cal{L}}(\ve{x},t)- \widetilde{\cal{L}}(\ve{x},t) ,      
\end{equation}
This Lagrangian produces the total Hamiltonian
\begin{eqnarray}
{\bo{\hat{H}}} &=&\int d^{3}x
\frac{1}{2m} \le ({\boldmath{\nabla}}{\psi}^{\a}{\boldmath{\nabla}}{\psi} 
- {\boldmath{\nabla}}{ {\widetilde{\psi}}}^{\a}{\boldmath{\nabla}} \widetilde{\psi} \ri )
\nonumber\\
&=& H-{ \widetilde{H}},
\label{rev-tot-ham-psi}
\end{eqnarray}
from which the Heisenberg equations can be obtained for both type of fields
\begin{equation}
i\dot{\psi}=[{\psi},{\hat{H}}]=[{\psi},{H}]\ \ \ , \ \ \ i{ \dot{\widetilde{\psi}}}=[ \widetilde{\psi},{\hat{H}}]=-[ \widetilde{\psi},\widetilde{H}].
\label{rev-Heis-psi}
\end{equation}
By solving the Heisenberg equations in cube of volume $V$ on obtains the expansion of the fields in terms of creation and annihilation operators
\begin{eqnarray} 
\psi (\ve{x},t) &=&\frac{1}{\sqrt{V}} 
\sum_{\ve{k}}e^{i \ve{k} \cdot \ve{x}}e^{-i\o_{\ve{k}}t}a_{\ve{k}}\ ,
\label{rev-exp-psi1}\\
\widetilde{\psi} (\ve{x},t)&=& \frac{1}{\sqrt{V}} 
\sum_{\ve{k}}e^{-i \ve{k} \cdot \ve{x}}e^{i\o_{\ve{k}}t} \widetilde{a}_{\ve{k}}\ ,
\label{rev-exp-psi2}\\
{\psi}^{\a} (\ve{x},t)&=& \frac{1}{\sqrt{V}} 
\sum_{\ve{k}}e^{-i \ve{k} \cdot \ve{x}}e^{i\o_{\ve{k}}t}a^{\a}_{\ve{k}}\ ,
\label{rev-exp-psi3}\\
{\widetilde{\psi}}^{\a} (\ve{x},t)&=& \frac{1}{\sqrt{V}} 
\sum_{\ve{k}}e^{i \ve{k} \cdot \ve{x}}e^{-i\o_{\ve{k}}t} \widetilde{a}^{\a}_{\ve{k}},
\label{rev-exp-psi4}
\end{eqnarray}
where the dispersion relation is ${\o_{\ve{k}}}=\frac{|\ve{k}|^{2}}{2m}$. By substituting the above equations into the definition of the total Hamiltonian given by the equation (\ref{rev-Heis-psi}), one obtains the total Hamiltonian for the excitations of the field and the tilde field
\begin{equation}
\hat{H}=\sum_{\ve{k}}\o_{\ve{k}}(a^{\a}_{\ve{k}}a_{\ve{k}}- \widetilde{a}^{\a}_{\ve{k}} \widetilde{a}_{\ve{k}}). 
\label{rev-HT}
\end{equation}
The total system is a collection of independent oscillators to which we can apply the TFD method. In general, the procedure of mapping each pair of oscillators and tilde oscillators from zero to finite temperature is not equivalent to the thermalization of the collection of all pairs simultaneously. This is due to the fact that the oscillators must obey the field structure and the symmetries of the original field and, also, the rules of the field interactions. However, in many cases of free fields, the two constructions are equivalent as a consequence of the independence of the oscillators.

\subsubsection{Thermalization of scalar field}

In order to thermalize the total system constructed above, one has to act with the Bogoliubov operator on the algebra of oscillators and on the total Hilbert space. Since the oscillators are free and independent, the Bogoliubov operator is given by the following equation
\begin{equation}
G=-i \sum_{\ve{k}} \th_{\ve{k}}(a^{\a}_{\ve{k}}  \widetilde{a}^{\a}_{\ve{\k}}
-\widetilde{a}_{\ve{\k}}a_{\ve{k}}).  
\label{rev-TB}
\end{equation}
The Bogoliubov transformation is generated by the unitary operator 
\begin{equation}
U(\th)=e^{-iG(\th)}=e^{- \sum_{\ve{k}} \th_{\ve{k}}(a^{\a}_{\ve{k}}  \widetilde{a}^{\a}_{\ve{k}}-\widetilde{a}_{\ve{k}}a_{\ve{k}})}.
\label{rev-Bog-tr-uni-psi}
\end{equation}
By acting with $U(\th)$ on the total vacuum which is the direct product of vacua of all field excitations, one obtains the thermal vacuum of the scalar field at finite temperature 
\begin{eqnarray}
|0(\b)\ra &=&U(\th)|0(\b)\ra \nonumber\\
&=&\prod_{\ve{k}} \le [{\cosh}^{-1}{\th}_{\ve{k}}(\b) \exp{({\tanh}{\th}_{\ve{k}}(\b){a}^{\a}_{\ve{k}} \widetilde{a}^{\a}_{\ve{k}})}\ri ]|0\ra \! \ra . 
\label{rev-vtc}
\end{eqnarray} 
Here, we have used the notation from the previous section
\begin{eqnarray}
u_{\ve{k}}(\b)&=&(1-e^{-\b \o_{\ve{k}}})^{-\frac{1}{2}}=\sqrt{1+f_{B}(\o_{\ve{k}})}=\cosh\th_{\ve{k}}(\b), \label{rev-uv}  \\
\v_{\ve{k}}(\b) &=&(e^{\b \o_{\ve{k}}}-1)^{-\frac{1}{2}}=\sqrt{f_{B}(\o_{\ve{k}})}=\sinh\th_{\ve{k}}(\b).  \label{rev-v}
\end{eqnarray}
The action of the Bogoliubov mapping (\ref{rev-Bog-tr-uni-psi}) on the oscillator operators is given by the following relations 
\begin{eqnarray}
{a}_{\ve{k}}(\b)&=&e^{-iG}a_{\ve{k}}e^{iG}={a}_{\ve{k}}\cosh{\th_{\ve{k}}}(\b)- \widetilde{a}^{\a}_{\ve{k}}\sinh{\th_{\ve{k}}}(\b), \label{rev-oac}\\
\widetilde{a}_{\ve{k}}(\b)&=&e^{-iG} \widetilde{a}_{\ve{k}}e^{iG}= \widetilde{a}_{\ve{k}}\cosh {\th_{\ve{k}}}(\b)-{a}^{\a}_{\ve{k}}\sinh {\th_{\ve{k}}}(\b)\ . \label{rev-oatc}
\end{eqnarray}
One can perform the inverse Bogoliubov transformation to express the oscillator operators at zero temperature in terms of operators at finite temperature and substitute these operators into the equation (\ref{rev-HT}). One obtains the following Hamiltonian at finite temperature
\begin{equation}
\hat{H}=\sum_{\ve{k}}\o_{\ve{k}} \le(a^{\a}_{\ve{k}}(\b)a_{\ve{k}}(\b)- \widetilde{a}^{\a}_{\ve{k}}(\b) \widetilde{a}_{\ve{k}}(\b) \ri).
\label{rev-HT-temp}
\end{equation} 
Alternatively, by computing the commutation relation between $G(\th)$ and $\hat{H}$ we find that $[ G(\th),\hat{H} ] = 0$ which guarantees that the thermalization of the scalar field is a process that is invariant dynamically. 

\subsubsection{Thermal Fock space of scalar field}

The field states in the Fock space can be constructed by acting with the thermal creation operators on the thermal vacuum given by the equation (\ref{rev-vtc}). The thermal vacuum satisfies the expected relations
\begin{eqnarray}
a_{\ve{k}}(\b)|0(\b)\ra &=& 0, 
\label{rev-va}\\
\widetilde{a}_{\ve{k}}(\b)|0(\b)\ra &=& 0.  
\label{rev-vat}
\end{eqnarray}
The one particle states are given by the following equations
\begin{eqnarray}
a^{\a}_{\ve{k}}(\b)|0(\b)\ra &=& e^{-iG}a^{\a}_{\ve{k}}|0\ra \! \ra,
\label{rev-one-part-psi1}\\
\widetilde{a}^{\a}_{\ve{k}}(\b)|0(\b)\ra &=&e^{-iG} \widetilde{a}^{\a}_{\ve{k}}|0\ra \! \ra.
\label{rev-one-part-psi2}
\end{eqnarray}
By successively applying the thermal creation operators, thermal states of different frequencies can be obtained as in the field theory at zero temperature. These states are eigenstates of the Hamiltonian from the equation (\ref{rev-HT-temp}). For example, one can easily show that $\hat{H}|0(\b)\ra = 0$. From that, it follows that
\begin{equation}
\la 0(\b)|H|0(\b)\ra =\la 0(\b)| \widetilde{H}|0(\b)\ra.
\label{rev-H-zero-psi}
\end{equation}
The number of thermal excitations is counted by the number operator $\hat{N}=N- \widetilde{N}$ which can be given in terms of operators at zero or finite temperature
\begin{eqnarray}
\hat{N} &=&\sum_{\ve{k}}(a^{\a}_{\ve{k}}a_{\ve{k}}- \widetilde{a}^{\a}_{\ve{k}} \widetilde{a}_{\ve{k}}), \label{rev-N-zero-psi}\\
\hat{N} &=&\sum_{\ve{k}}\le (a^{\a}_{\ve{k}}(\b)a_{\ve{k}}(\b)- \widetilde{a}^{\a}_{\ve{k}}(\b) \widetilde{a}_{\ve{k}}(\b)\ri ).   
\label{rev-N-temp-psi}
\end{eqnarray}
Its invariance under the Bogoliubov transformation is expected since $\hat{N}$ is related to $\hat{H}$ by the canonical relation
\begin{eqnarray}
\hat{H}&=&\sum_{\ve{k}}{\hat{N}}_{\ve{k}}\o_{\ve{k}},
\label{rev-HN-psi}\\
{\hat{N}}_{\ve{k}}&=&a^{\a}_{\ve{k}}(\b)a_{\ve{k}}(\b)- \widetilde{a}^{\a}_{\ve{k}}(\b) \widetilde{a}_{\ve{k}}(\b).
\label{rev-Nk-psi}
\end{eqnarray}
It follows from (\ref{rev-H-zero-psi}) and (\ref{rev-HN-psi}) that
\begin{equation}
\la 0(\b)|N|0(\b)\ra =\la 0(\b)| \widetilde{N}|0(\b)\ra . 
\label{eq-N-psi}
\end{equation}
This equality shows that there is the same number of excitations from the original field as well as from the reservoir field at any given temperature. If one computes
\begin{eqnarray}
a^{\a}_{\ve{k}}(\b)|0(\b)\ra &=&\frac{1}{\sinh{{\th}_{\ve{k}}}(\b)} \widetilde{a}_{\ve{k}}|0(\b)\ra \label{rev-buraco1},\\
\widetilde{a}^{\a}_{\ve{k}}(\b)|0(\b)\ra &=&\frac{1}{\sinh{{\th}_{\ve{k}}}(\b)}{a}_{\ve{k}}|0(\b)\ra \label{rev-buraco2},
\end{eqnarray}
one can see that creating a thermal excitation of a given frequency corresponds to annihilating a zero temperature excitation of the same frequency and vice versa. Due to this fact, the thermal excitation is 
sometimes interpreted as a physical hole \cite{hu}.

\subsubsection{Kubo-Martin-Schwinger conditions} 

As we have reviewed in the previous sections, in the the TFD one can compute consistently the statistical averages the field observables. Moreover, the thermal field is a quantum field theory in which the standard computational methods can be straightforwardly applied. It can be shown that the transition amplitudes can be expressed in terms of thermal Green functions. However, in order to compute them, one needs to impose boundary conditions consistent with the thermal equilibrium. 
These are known as the {\em Kubo-Martin-Schwinger (KMS) conditions}. For two arbitrary observables, they are defined along the time-like direction the following identities
\begin{eqnarray}
\la 0(\b)|A(t)B(t')|0(\b) \ra  &=& \la 0(\b)|\w{A}^{\a}(t)e^{\b \hat{H}/2}B(t')|0(\b) \ra \non
&=& \la 0(\b)|\w{A}^{\a}(t+i\b/2)B(t')|0(\b) \ra \non
&=& \la 0(\b)|B(t')e^{\b \hat{H}/2}\w{A}^{\a}(t+i\b/2)|0(\b) \ra \non
&=& \la 0(\b)|B(t')A(t+i\b)|0(\b) \ra . \label{rev-kms}
\end{eqnarray}
The above relations represent prescriptions for calculating the thermal propagators. They can be obtained from the equations (\ref{rev-vou}) and (\ref{rev-vou1}).
      
\subsection{Entropy and free energy}

Beside the usual observables that are defined in the Quantum Field Theoy at zero temperature, the fields at finite temperature have observables associated with the statistical properties of the ensembles of string excitations. In the TFD formalism it is possible to define these thermal observables. We are going to recall the definition of two of them: the entropy operator and the free energy operator in the case of the free bosonic field discussed in the previous section.

\subsubsection{Entropy operator}

The entropy operator is defined such that it reproduces the entropy in the grand canonical ensemble
\begin{equation}
S=k_{B}\sum_{{k}}\le \{ (1+ \langle n_{{k}}\ra )\ln(1+ \langle n_{{k}}\ra )- \langle n_{{k}}\ra \ln  \langle n_{{k}}\rangle \ri \},
\label{rev-S-op}
\end{equation}
where $n_{k}$ is the average ocupation number of the $k$ state. According to the fundamental hypothesis of the TFD formalism, the statistical entropy $S$ should be the vacuum expectation value of the entropy operator $K$
\begin{equation}
S=k_{B} \langle K\ra \equiv k_{B} \langle 0(\b)|K|0(\b)\ra ,
\label{rev-K-op-def}
\end{equation}
where $k_B$ is the Boltzmann constant. This equation is satisfied by the operator 
\begin{equation}
K=-\sum_{\ve{k}} \le(a^{\a}_{\ve{k}}a_{\ve{k}}\ln{\sinh^{2}{\th_{\ve{k}}(\b)}}-a_{\ve{k}}a^{\a}_{\ve{k}}\ln{\cosh^{2}{\th_{\ve{k}}(\b)}}\ri ). 
\label{rev-centropia}
\end{equation}
To it, one can associate the operator $\tilde{K}$ that is obtained by substituting the oscillators by tilde-oscillators. The operators $K$ and $\tilde{K}$ factorize the unitary Bogoliubov transformation
$U(\th)$ given by the equations (\ref{rev-Bog-tr-uni-psi}) and (\ref{rev-vtc}) as following
\begin{equation}
|0(\b)\ra =e^{-K/2} \le \{\exp{\sum_{\ve{k}}a^{\a}_{\ve{k}} \widetilde{a}^{\a}_{\ve{k}}}\ri \}|0\ra \! \ra =e^{- \widetilde{K}/2}\le \{ \exp{\sum_{\ve{k}} \widetilde{a}^{\a}_{\ve{k}}{a}^{\a}_{\ve{k}}} \ri \}|0\ra \! \ra . 
\label{rev-vtk}
\end{equation}
The entropy operators $K$ and $\tilde{K}$ define new mappings from the coherent state $|\hat{I}\ra$ to the thermal vacuum
\begin{eqnarray}
|0(\b)\ra &=& e^{-K/2}|\hat{I}\ra =e^{- \widetilde{K}/2}|\hat{I}\ra ,
\label{rev-I-map}\\
|\hat{I}\ra &=&\sum_{n}|n, \widetilde{n}\ra = \exp \le \{\sum_{\ve{k}}a^{\a}_{\ve{k}} \widetilde{a}^{\a}_{\ve{k}}\ri \}|0\ra \! \ra.
\label{rev-ent-n-tilde-n}
\end{eqnarray}
The total entropy operator $\hat{K}= K- \widetilde{K}$ commutes with the Bogoliubov operator given by the equation (\ref{rev-TB}): $[K- \widetilde{K},G]=0$. By acting with $\hat{K}$ on the thermal vacuum one obtains
\begin{equation} 
(K- \widetilde{K})|0(\b)\ra = 0. \label{rev-K-tilde-K}
\end{equation}
This equation shows that the same value of the entropy $S$ is obtained from either $K$ or $\tilde{K}$. 
The entropy operator can be used to fix the value of the functions $\th_{\ve{k}}(\b)$ to 
\begin{equation} 
\th_{\ve{k}}(\b) = \mbox{arcsinh}(\sqrt{n_{\ve{k}}}). 
\label{rev-nk}
\end{equation}
The equation equation (\ref{rev-nk}) can be obtained from the equations (\ref{rev-S-op}) and (\ref{rev-K-op-def}) and by assuming that the vacuum expectation value in the thermal vacuum of the total Hamiltonian $\hat{H}$ is constant \cite{hu}.

\subsubsection{Free energy}

The Helmholtz free energy is by definition the following thermodynamical function
\begin{equation}        
\O = -TS+ E -\mu \mathcal{N},
\label{rev-Helmholtz-def}
\end{equation}
where $E$ is the average energy of the system and $\mathcal{N}$ is the average particle number. By substituting the vacuum expectation values of $K$ and $N$ in these average values, one obtains the following expression for the free energy of a system of oscillators
\begin{eqnarray}
\O &=& -TS+ \la H\ra -\mu \la N\ra \nonumber\\
&=&\sum_{{k}}
\left[-\fr{1}{\b}\left( (1+n_{\ve{k}})\ln(1+n_{\ve{k}})-n_{\ve{k}}\ln n_{\ve{k}}\right) + (\e_{\ve{k}}-\mu)n_{\ve{k}}  
\right].
\label{rev-elh} 
\end{eqnarray}
Let us analyze the thermal equilibrium from the point of view of the free energy. In this case, the average energy is constant and the arbitrary variation of the free energy around the equilibrium point is zero $\d \O = 0$. It follows that $ {\partial \O} / {\partial n_{\ve{k}}}=0$ from which one obtains
\begin{equation}
\th_{{\ve{k}}} = \mbox{arcsinh} 
\left[ \frac{e^{-\frac{1}{2}\b (\o_{\ve{k}}-\b)}}{\sqrt{1-e^{-\frac{1}{2}\b (\o_{\ve{k}}-\b)}}}\right].
\label{rev-th-2}
\end{equation}
This is the same value of the function $\th_{{\ve{k}}}$ as the one obtained in the equation (\ref{rev-nk}) but with the average number of ${\ve{k}}$ excitations expressed in terms of the Bose-Einstein distribution.

\subsection{Axioms of Thermo Field Dynamics}

It is possible to derive a set of axioms for the Thermo Field Dynamics from the above construction that are useful for the generalization of the formalism to other systems and for applications \cite{hu}. Let $\Im=\{A\}$  and $\w{\Im}=\{\w{A}\}$ be two independent algebras of operators. 

\subsubsection{General axioms}

{\em Axiom 1} Two variables that belong to the independent algebras $A \in \Im \ \mbox{e} \ \w{B} \in \w{\Im} $ are independent 
\begin{equation}
[A,\w{B}]=0.
\label{rev-axiom1}
\end{equation}

{\em Axiom 2} There is an internal operation $\tilde{\ }$ of the direct product algebra $\Im \otimes \w{\Im}$ such that
\begin{eqnarray}
& \mbox{a)}& \ \w{(AB)}=\w{A}\w{B},  
\label{rev-axiom2-a}\\ 
& \mbox{b)} & \ \w{{(c_{1}A+c_{2}B)}}=c^{\ast}_{1}\w{A}+c^{\ast}_{2}\w{B}, 
\label{rev-axiom2-b}\\
& \mbox{c)}& \ \w{A^{\a}}=\w{A}^{\a},
\label{rev-axiom2-c}
\end{eqnarray}
for any $A, B \in \Im $ and any $\w{A} , \w{B} \in \w{\Im}$. The complex numbers $c_{1}$ and $c_{2}$ are arbitrary.
 
{\em Axiom 3}  The thermal vacuum is defined by the following {\em thermal state conditions}
\begin{eqnarray}
A(t,\ve{x})|0(\b) \ra=\s \w{A}^{\a}(t-i\b/2, \ve{x})|0(\b) \ra, \label{rev-cet1}
\\
\la 0(\b)|A(t,\ve{x})=\la 0(\b)| \w{A}^{\a}(t+i\b/2), \ve{x}) \s^{\ast}, \label{rev-cet2}
\end{eqnarray} 
where $\s = 1$ for any bosonic operator $A \in \Im$.

{\em Axiom 4} The thermal vacuum is invariant under the tilde conjugation
\begin{equation}
\w{|0(\b) \ra}=|0(\b) \ra.
\label{rev-axiom3}
\end{equation} 

{\em Axiom 5} The energy-momentum operator $P_{\mu} \in \Im $ generates the space-time translations in the subalgebra $\Im $ by the following action
\begin{equation}
A(x)=e^{iP_{\mu}x^{\mu}}Ae^{-iP_{\mu}x^{\mu}},
\label{rev-axiom4}
\end{equation}
for any $A \in \Im$.

{\em Axiom 6 }. The tilde conjugation is an involution for bosons with
\begin{equation}
\w{\w{A}}= \s A,
\label{rev-axiom6}
\end{equation}
for any $A \in \Im$. 

The first axiom can be generalized by using the second and the fourth one to the statement that 
if $A(x) \in \Im \ $ and $ \w{B}(y) \in \w{\Im} $ are arbitrary operators, then they satisfy 
\begin{equation} 
[A(x),\w{B}(y)]=0,
\label{rev-axiom1-gen}
\end{equation}
where $x$ and $y$ are arbitrary space-time events.

\subsubsection{Thermal doublet formalism from axioms}

The second axioms allows one to generalize the Heisenberg equations and the commutation relations from the Quantum Field Theory to the TFD formalism. Also, from this axiom one can derive a combination of operators $A(x) \in \Im $ and $ \w{B}(y) \in \w{\Im} $ that annihilate the thermal vacuum. From the second axiom, the sign of the combination of the hermitian conjugation and the tilde conjugation in any order is unchanged. This allows one to define the {\em thermal doublet formalism} as follows. The thermal doublet operator is defined by the relations
\begin{equation}
A^{\al}= \cases {A, \ \mbox{if} \ \al=1 \cr
\w{A}^{\a}, \ \mbox{if} \ \al=2.  \cr }   
\label{rev-dt}
\end{equation} 
One can generalize the commutator relations $[A(x),B(y)]=C(x,y)$ to the algebra of the thermal doublet operators 
\begin{equation}
[A^{\al}(x),B^{\al}(y)]=\t^{\al \g}C^{\g}(x,y),
\end{equation}
where
\begin{equation}
\t=\fourmat{1}{0}{0}{-1} 
\label{rev-tau}.
\end{equation}
From the definition (\ref{rev-dt}), one can cast an arbitrary operator $F(A)$ in the thermal doublet form by using the following equation
\begin{equation}
[F(A)]^{\al}=P_{\al}\left[ F(A^{\al}) \right], \label{rev-oter}
\end{equation}
where $P_{\al}$ is the {\em thermal ordering operator} defined by the relation 
\begin{equation} 
P_{\al}[A^{\al}B^{\al}\dots C^{\al}]=\cases {A^{1}B^{1} \dots C^{1}, \ \mbox{if} \ \al=1 , \cr
 C^{2} \dots B^{2}A^{2}, \ \mbox{if} \ \al=2. \cr} \label{rev-operador}
\end{equation}
The Heisenberg equations take the following form in the thermal doublet notation
\begin{equation} 
i\partial_{\mu}\ps^{\al}(x)=\e^{\al}[\ps^{\al}(x),P^{\al}_{\mu}],
\hspace{0.5cm} 
\e^{\al}= \cases {1, \ \mbox{if} \ \al=1   \cr
                 -1, \ \mbox{if} \ \al=2.  \cr }
\label{rev-eht}
\end{equation}                             
By using the generator of the space-time translations
\begin{equation}
\hat{P}_{\mu}=\sum_{\al}\e^{\al}P^{\al}_{\mu}=P_{\mu}-{\w{P}}_{\mu},
\end{equation}
we can put the Heisenberg equations into the following form
\begin{equation} 
i\partial_{\mu}\ps^{\al}(x)=[\ps^{\al}(x),\hat{P}_{\mu}]. 
\label{rev-ehts}
\end{equation}
The thermal Lagrangian (Hamiltonian) can be obtained from the Lagrangian (Hamiltonian) of any system 
as follows  
\begin{eqnarray}
\hat{L}=\sum_{\al}\e^{\al}L^{\al}=L-\w{L},
\label{rev-Lag-doublet}\\
\hat{H}=\sum_{\al}\e^{\al}H^{\al}=H-\w{H}, 
\label{rev-Ham-doublet}
\end{eqnarray}

The third axiom allows one to construct annihilation and creation operators for the thermal vacuum from any non-Hermitian operator by applying the relations
\begin{equation}
A(t)|0(\b)\ra= {\w{A}}^{\a}(t-i\b/2)|0(\b)\ra,
\label{rev-th-vac-A1}
A^{\a}(t)|0(\b)\ra={\w{A}}(t-i\b/2)|0(\b)\ra.
\label{rev-th-vac-A2}
\end{equation}

From the above axioms from the equations (\ref{rev-axiom1}) - (\ref{rev-axiom6}) one can derive all the previous elements of the TFD formalism, such as the Bogoliubov operators, the KMS conditions, etc. For more details we reefer the reader to \cite{hu,mnu}.

\section{Strings at Finite Temperature}

In this section, we are going to review the results obtained by applying the TFD method to some simple configurations of bosonic strings and $D$-branes. Our presentation follows mainly \cite{Vancea:2000gr,Abdalla:2001ad,Abdalla:2002ak} and
\cite{PaniagodeSouza:2002nz,Graca:2007bx}.

\subsection{Thermalization of open string}

The bosonic string in the Minkowski space-time and in the conformal gauge was presented in the Section 2. The light-cone quantization preserved only the physical states. Since the degrees of freedom form a system of bosonic fields on the two dimensional world-sheet, one can study their properties at finite temperature by using the TFD method presented in the Section 3.

Consider the bosonic open string with boundary conditions given by the equations (\ref{rev-Neumann-bc}) and (\ref{rev-Dirichlet-bc}). The solutions of the equations of motion with different combinations of boundary conditions were given in the equations (\ref{rev-NN-bc})-(\ref{rev-DN-bc}). After applying the canonical quantization in the light-cone gauge
(\ref{rev-X-plus})-(\ref{rev-light-cone-st-01}), the Fourier modes were interpreted as operators that characterized the quantum string excitations. It is useful to rescale these operators to obtain the canonical oscillator operators
\be
A^{\m}_n = \frac{1}{\sqrt{n}}\alpha^{\m}_n , \hspace{0.5cm} A^{\m \dagger}_n = 
\frac{1}{\sqrt{n}}\alpha^{\m}_{-n}~~,
~~~ n > 0,
\label{rev-oscop}
\ee 
where $\mu = 1, \ldots, 24$ in the light-cone gauge. The operators (\ref{rev-oscop}) satisfy the canonical commutation relations among themselves. Also, they commute with the coordinates and momenta of the center of mass. Their action on the string vacuum is given by the following relations
\bea
A^{\m}_n \left| 0 \right\rangle ~ &=& ~ 0 ~~,~~~ \forall n, \label{rev-vacosc}\\
\hat{p}^{\m} \left| p \right\rangle ~ & = & ~ p^{\m} \left| p \right\rangle 
\label{rev-vacmom}.
\eea
The TFD method can be applied to obtain the thermal bosonic string. The first step is to duplicate the string by enlarging the system with the degrees of freedom of the thermal reservoir denoted by $~\tilde{}~$. The total system has the Hilbert space
\be
\hat{\cal H}={\cal H} \bigotimes \w{\cal H} .
\label{rev-total-Hilbert-string}
\ee
The unitary and tilde invariant Bogoliubov operator can be defined for 
each oscillator $n$ in each direction $\m$ by the following relation
\be
G^{\m}_n ~=~-i \th_n(\beta_T)(A_n \cdot \tilde{A}_n - \tilde{A}_n^{\dagger} \cdot
 A_n^{\dagger} ).
\label{rev-bogolop}
\ee  
Here, $\th_n(\beta_T)$ parameter is fixed by the equation (\ref{rev-nk}) with $\ve{k}$ replaced by $n$. The dot in (\ref{rev-bogolop}) represents the Euclidean scalar product in the transverse target space. The Bogoliubov generator maps the total string vacuum at zero temperature to the thermal vacuum as in the equations (\ref{rev-tbv}) and (\ref{rev-vtc})
\be
\left. \left| 0(\beta_T ) \right\rangle \! \right\rangle ~= ~\prod_{ m > 0} 
e^{-iG_m} 
\left. \left| 0 \right\rangle \! \right\rangle ,
\label{rev-vacT}
\ee
where $\left. \left| 0 \right\rangle \! \right\rangle= 
\left| 0 \right\rangle\tilde{\left| 0 \right\rangle} $ and
\be
G_n = \sum_{\m=1}^{24}G^{\m}_{n}.
\label{rev-opbogoltot}
\ee
The thermal vacuum of the string has the following form 
i.e.
\be
\left. \left| \O (\beta_T) \right\rangle \! \right\rangle ~=~ 
\left. \left| 0(\beta_T) \right\rangle \! \right\rangle
\left| p \right\rangle \left| \tilde{p} \right\rangle. 
\label{rev-totvacT}
\ee
The state (\ref{rev-vacT}) is annihilated by 
all thermal annihilation operators. These can be constructed as in the equations
(\ref{rev-aa}) -(\ref{rev-atd1}) 
\be
A^{\m}_{n}(\beta_T) ~= ~ e^{-iG_n}A ^{\m}_{n}e^{iG_n}~~~,~~~
\tilde{A}^{\m}_{n}(\beta_T) ~= ~ e^{-iG_n}\tilde{A} ^{\m}_{n}e^{iG_n},
\label{rev-annihT}
\ee
with the corresponding equations for the creation operators 
$A^{\m \dagger}_{n}(\beta_T)$ and $\tilde{A}^{\m \dagger}_{n}(\beta_T)$.
We note that the coordinates and the momenta of the center of mass of string are 
invariant under the Bogoliubov mapping. 

From the above considerations, we conclude that the thermal string solution $X^{\m}(\beta_T)$ 
at $T \neq 0$ can be obtained by replacing the operators from the equations (\ref{rev-NN-bc})-(\ref{rev-DN-bc}) 
with the corresponding operators at $T \neq 0$. This is the result of the action of the Bogoliubov 
transformations on  the string operators $X^{\mu}$ at $T = 0$.

The generators of the Virasoro algebra at $T \neq 0$ can be obtained by acting with 
the Bogoliubov operators on the Virasoro generators at $T=0$ or by assembling them from 
the thermal oscillator operators. Therefore, the Virasoro algebra closes on the thermal string
states. 
The Bogoliubov transformation maps the two string copies from the total string into two thermal string 
copies. The relation between these systems is complicated since the two strings at zero temperature get mixed at higher temperatures. From the invariance of the thermal vacuum to the tilde involution, it follows that the tilde and non-tilde excitations are generated simultaneously at finite temperature. 

\subsection{Entropy of string fields}

It is interesting to see what is the entropy of the thermal open string fields $X^{\mu}(\beta_T)$ with different boundary conditions given in the equations (\ref{rev-NN-bc})-(\ref{rev-DN-bc}). The entropy operator is defined in the equation (\ref{rev-K-op-def}) with $K$ given by the equation (\ref{rev-centropia}). By using these equations, we can write down the entropy operators $K$ and $\tilde{K}$ for the open string as
\bea
K ~&=& ~\sum_{\m = 1}^{24}\sum_{n=1}^{\infty}( A^{\m \dagger}_n A^{\m}_n 
\log \sinh^2 \th_n -A^{\m}_n A^{\m \dagger}_n \log \cosh^2 \th_n ),
\label{rev-entropy}\\
\tilde{K} ~&=&  \sum_{\m = 1}^{24}\sum_{n=1}^{\infty}(\tilde{A}^{\m \dagger}_n
\tilde{A}^{\m}_n \log \sinh^2 \th_n -
\tilde{A}^{\m}_n \tilde{A}^{\m \dagger}_n \log \cosh^2 \th_n ).
\label{rev-entropytilde}
\eea
In order to compute $K$, one can factorize it according to the space-time directions
\be
K ~ =~ \sum_{\m = 1}^{24} K^{\m}.
\label{rev-decomentrop}
\ee

Consider the string fields with the NN boundary conditions. The entropy of the thermal string fields in this state is given by the expectation value of $K$ operator 
$\left\langle\!\left\langle X^{\m}(\beta_T)\left| K^{\rho} \right| X^{\m}(\beta_T )
\right\rangle\!\right\rangle$. One can compute it by evaluating the center of mass and the oscillator contributions separately. 
\bea
\left\langle \!\left\langle X^{\mu}(\beta_T)\left| K^{\rho} \right| 
X^{\m}(\beta_T )\right\rangle\!\right\rangle
~&=&~{\mbox{CM terms}} \nonumber\\
&-& 2\alpha ' \sum_{n,k,l >0}\frac{e^{i(l-n)\t}}{\sqrt{ln}}\cos n\s \cos l\s
\left[ (T_1)^{\m\rho\n}_{nkl} + (T_2)^{\m\rho\n}_{nkl}\right].
\label{rev-matrentr}
\eea
The terms from the oscillator contribution have the following form
\bea
(T_1)^{\m\rho\n}_{nkl} ~&=&~
\left\langle \tilde{0} \left| \left\langle 1^{\m}_n \left|
\prod_{m>0}e^{-iG_m}A^{\rho \dagger}_{k}A^{\rho}_{k}\log \sinh^2 \theta_k 
\prod_{s>0}e^{iG_s}
\right| 1^{\n}_l \right\rangle \right| \tilde{0} \right\rangle
\left\langle \tilde{p} \left|  \tilde{q} \right\rangle \right.
\left\langle p \left| q \right \rangle \right. ,\nonumber\\
(T_2)^{\m\rho\n}_{nkl} ~&=&~
- \left\langle \tilde{0} \left| \left\langle 1^{\m}_n \left|
\prod_{m>0}e^{-iG_m}A^{\rho }_{k}A^{\rho \dagger}_{k}\log \cosh^2 \theta_k 
\prod_{s>0}e^{iG_s}
\right| 1^{\n}_l \right\rangle \right| \tilde{0} \right\rangle
\left\langle \tilde{p} \left|  \tilde{q} \right\rangle \right.
\left\langle p \left| q \right \rangle \right.
\label{rev-Ts}
\eea
We have used the following notation for the one-excitation state
\be
\left| 1^{\m}_{l} \right\rangle ~= ~A^{\m \dagger}_{l} \left| 0 \right\rangle.
\label{rev-field}
\ee
In order to compute the matrix elements, one needs to normalize the momentum states in the transverse space. We use the finite volume $V_{24}$ normalization  
\bea
\left.\left\langle p \right| q \right\rangle &~=~& 2 \pi 
\delta^{(24)} (p-q),
\label{rev-normstate1}\\
(2\pi )^{24}\delta^{(24)}(0) &~=~& V_{24}.
\label{rev-normstate2}
\eea
The result of the calculation of the oscillator contribution to the matrix element of the entropy operator is given by the following equation
\bea
\left\langle\!\left\langle X^{\m}(\beta_T)\left| K^{\rho} \right| X^{\m}(\beta_T )
\right\rangle\!\right\rangle
~&=&~{\mbox{CM terms}} 
-2 \alpha ' (2 \pi)^{(48)} \delta^{\m \n} \delta^{(24)}(p-q)\delta^{(24)}
(\tilde{p}-\tilde{q}) \times 
\nonumber\\
& &\sum_{n>0}\frac{1}{n}\cos^2 n\s[\log (\tanh \th_n )^2\delta^{\rho \n } -
\delta^{\rho \rho }\sum_{k>0} \delta_{k k}].
\label{rev-matrelemCM}
\eea
In order to calculate the contribution of the center of mass, one has to normalize the coordinate-momenta matrix. The normalization used is
\be
\left. \left\langle x \right| p \right\rangle ~=~(2\pi \hbar ) ^{-12}
e^{i p \cdot x / \hbar}.
\label{rev-matrixxp}
\ee
A simple algebra gives the final result
\bea
& &\left\langle \!\left\langle X^{\mu}(\beta_T)\left| K^{\rho} \right| 
X^{\m}(\beta_T )\right\rangle\!\right\rangle
~=~ \nonumber\\
&- &(2\pi \hbar)^{-24}\left[ (2\pi\hbar)^{24}(2\alpha ' \t)^2 p^{\m} p'^{\n}
\delta^{(24)}(p-p') 
+ 
2\alpha ' \t (I^{\m}_2p'^{\n} + I'^{\n}_2 p^{\m}) + 
I^{\m}_2I^{\n}_2\prod_{j \neq \m , \n}I^j_1 \right]\nonumber\\ 
&\times &\delta^{(24)}(\tilde{p} - \tilde{p'})\sum_{m=1} 
\left[{\mbox n}^{\rho}_m 
\log {\mbox  n}^{\rho}_m + (1- {\mbox  n}^{\rho}_m) 
\log( 1- {\mbox  n}^{\rho}_m ) 
\right]
- 2 \alpha ' (2 \pi)^{(48)} \delta^{\m \n} \delta^{(24)}(p-p')\nonumber\\
&\times &\delta^{(24)}(\tilde{p}-\tilde{p'})  
\sum_{n>0}\frac{1}{n}\cos^2 n\s \left[ \log (\tanh \th_n )^2\delta^{\rho \n } -
\delta^{\rho \rho }\sum_{k>0} \delta_{k k} \right],
\label{rev-entropyRho}
\eea
The entropy is expressed in terms of unidimensional integrals on the finite domains 
$ x \in \left[ x_0, x_1\right] $ that are given by
\bea
I_1 &~=~& -i \hbar (p'-p)^{-1}\left[ e^{\frac{i}{\hbar}(p'-p)x_1} - 
e^{\frac{i}{\hbar}(p'-p)x_0}\right],
\label{rev-int1}\\
I_2 &~=~& -i \hbar (p'-p)^{-1}\left[  -i\hbar I_1 + 
x_1e^{\frac{i}{\hbar}(p'-p)x_1} 
- x_0 e^{\frac{i}{\hbar}(p'-p)x_0}\right].
\label{rev-int2}
\eea 
In the above equations, $\left| p \right\rangle$ and
$\left| p' \right\rangle$ are the momenta of the final and initial states, respectively, and 
\be
{\mbox n}^{\rho}_{m} ~=~ \left\langle \!\left\langle 0(\beta_T) \left| 
A^{\rho \dagger}_m A^{\rho}_m \right| 0(\beta_T) \right\rangle\!\right\rangle 
= \sinh^2 \th_m \label{rev-Nnumber} 
\ee 
represents the number of string excitations in the thermal vacuum. The natural interpretation of the result obtained in the equation (\ref{rev-entropyRho}) is that of the entropy of the thermal open string excitations in states with 
NN boundary conditions. Similar computations can be performed to calculate the matrix elements of $K$ operator between states with DD, DN and ND boundary conditions. The non-vanishing matrix elements are given by the following relations
\bea 
{\mbox DD}&:& 
\left\langle \!\left\langle X^{\mu}(\beta_T)\left| K^{\rho} \right| 
X^{\m}(\beta_T )\right\rangle\!\right\rangle
=
2 \alpha ' (2 \pi)^{(48)} \delta^{\m \n} \delta^{(24)}(p-p')\delta^{(24)}
(\tilde{p}-\tilde{p'})  
\nonumber\\
&\times &
\sum_{n>0}\frac{1}{n}\sin^2 n\s \left[ \log (\tanh \th_n)^2\delta^{\rho \n } -
\delta^{\rho \rho }\sum_{k>0} \delta_{k k} \right],
\label{rev-entrDD}\\
{\mbox DN}&:&
\left\langle \!\left\langle X^{\mu}(\beta_T)\left| K^{\rho} \right| 
X^{\m}(\beta_T )\right\rangle\!\right\rangle
= 
2 \alpha ' (2 \pi)^{(48)} \delta^{\m \n} \delta^{(24)}(p-p')\delta^{(24)}
(\tilde{p}-\tilde{p'})\!  
\nonumber\\
&\times &
\sum_{r=\ZZ + 1/2 }\!\frac{1}{r}\sin^2 r\s \left[ \log (\tanh \th_r)^2
\delta^{\rho \n } -
\delta^{\rho \rho }\sum_{k>0} \delta_{k k} \right],
\label{rev-entrDN}\\
{\mbox ND}&:&
\left\langle \!\left\langle X^{\mu}(\beta_T)\left| K^{\rho} \right| 
X^{\m}(\beta_T )\right\rangle\!\right\rangle
= 
2 \alpha ' (2 \pi)^{(48)} \delta^{\m \n} \delta^{(24)}(p-p')
\delta^{(24)}(\tilde{p}-\tilde{p'}) 
\nonumber\\
&\times & \!\sum_{r=\ZZ + 1/2}\!\frac{1}{r}\cos^2 r\s \left[ \log (\tanh \th_r)^2
\delta^{\rho \n } -
\delta^{\rho \rho }\sum_{k>0} \delta_{k k}, \right]
\label{rev-entrND}
\eea
where $\ZZ + 1/2$ are half-integer numbers. The contribution of just a 
single field is obtained by taking $\m = \rho = \n$.
The entropy given by the equations (\ref{rev-entropyRho}), (\ref{rev-entrDD}), (\ref{rev-entrDN}) and 
(\ref{rev-entrND}) is the entropy of states associated to the general
solutions of the equations of motion. This is not the entropy of the thermal string calculated in the thermal vacuum. Indeed, the later is the entropy of the string oscillators in all directions and should not depend on the boundary condition, while the former is a function of the world-sheet.

\subsection{Thermalization of closed string}

The thermalization of the closed string follows the same line as of the open string presented above. The canonical oscillator operators are defined by the normalization relations
\be
A^{\m}_n = \frac{1}{\sqrt{n}}\al^{\m}_n ~~~;~~~A^{\m \dagger}_n = 
\frac{1}{\sqrt{n}}\al^{\m}_{-n}~~,
\ee
\be
B^{\m}_n = \frac{1}{\sqrt{n}}\b^{\m}_n ~~~;~~~B^{\m \dagger}_n = 
\frac{1}{\sqrt{n}}\b^{\m}_{-n}~~,
\ee
for all $n>0$. The total string at zero temperature is obtain by duplicating the above oscillators by adding the tilde oscillators
\be
\w{A}^{\m}_n = \frac{1}{\sqrt{n}}\w{\al}^{\m}_n ~~~;~~~\w{A}^{\m \dagger}_n = 
\frac{1}{\sqrt{n}}\w{\al}^{\m}_{-n}~~,
\ee
\be
\w{B}^{\m}_n = \frac{1}{\sqrt{n}}\w{\b}^{\m}_n ~~~;~~~\w{B}^{\m \dagger}_n = 
\frac{1}{\sqrt{n}}\w{\b}^{\m}_{-n}.
\ee
The algebra of the oscillator operators is given by the following relations
\be
[A^{\m}_n,A^{\nu \a}_m]=[\w{A}^{\m}_n,\w{A}^{\nu \a}_m]=\d_{n+m} \h^{\mu \nu} \ \ , \ \ [B^{\m}_n,B^{\nu \a}_m]=[\w{B}^{\m}_n,\w{B}^{\nu \a}_m]=\d_{n+m} \h^{\mu \nu}, \label{comut1}
\ee
\be
[A^{\m}_{n},{\w{A}}^{\nu}_{m}]=[A^{\m}_n,{\w{A}}^{\nu \a}_m]=[A^{\m}_{n},{\w{B}}^{\nu}_{m}]= \cdots =0.\label{comut2}
\ee
The total Fock space is the direct product of the string and tilde string Fock spaces
\be
\hat{\cal H}={\cal H} \bigotimes \w{\cal H} .
\nonumber
\ee
Since each Fock space is already a direct product of spaces corresponding to the left- and right-moving modes, one has to introduce a notation for the vectors from the total space. We denote by $ | ~~ \ra \ra $ an arbitrary vector from $\hat{\cal H}$. The vacua of string oscillators in each sector have the following form 
\be 
{|0 \ra \ra} _{\al}={|0 \ra}_{\al} \bigotimes \w{|0 \ra}_{\al}=|0,0 \ra_{\al}, \hspace{0.25cm} {|0 \ra \ra} _{\b}={|0 \ra}_{\b} \bigotimes \w{|0 \ra}_{\b}=|0,0 \ra_{\b}. \label{etca}
\ee
Then the total vacuum of the oscillators is the following direct product 
\bea
|0 \ra \ra={|0 \ra \ra} _{\al}{|0 \ra \ra} _{\b}= \le ({|0 \ra}_{\al} \bigotimes \w{|0 \ra}_{\al} \ri )\le ({|0 \ra}_{\b} \bigotimes \w{|0 \ra}_{\b} \ri ) \non =\le ({|0 \ra}_{\al} \bigotimes {|0 \ra}_{\b} \ri )\le (\w{|0 \ra}_{\al} \bigotimes \w{|0 \ra}_{\b} \ri ). \label{etcf}
\eea
Note that the last equation is the result of the independence of the degrees of freedom of the string and the thermal reservoir. The total vacuum of the string is obtained by multiplying the state from (\ref{etcf}) by  $ |p \ra \bigotimes \w{|p \ra }$. 

As described in the Section 3, the thermal string can be obtained from the total string by applying the unitary and tilde invariant Bogoliubov transformations. In the oscillator sector of the total string, the Bogoliubov operators are defined as
\bea 
G^{\al}_{n} &=& -i\th(\b_{T})\le (A_{n} \cdot {\w{A}}_{n}-A^{\a}_{n} \cdot {\w{A}}^{\a}_{n} \ri ), \non
G^{\b}_{n}&=& -i\th(\b_{T})\le (B_{n} \cdot {\w{B}}_{n}-B^{\a}_{n} \cdot {\w{B}}^{\a}_{n} \ri ). \label{obtf}
\eea
The $\th_{n}(\b _{T})$ function is the same for the oscillators $n$ from the left- and right-moving sectors. The Bogoliubov operators are Hermitian and satisfy the following standard TFD algebra
$$
[G_{n}^{\al},A^{\mu}_{n}]=-i \th_{n}({\b}_{T}){\w{A}}_{n}^{\mu \a}, \ \ \ \  [G_{n}^{\al},B^{\mu}_{n}]=-i \th_{n}({\b}_{T}){\w{B}}_{n}^{\mu \a}, 
$$
$$
[G_{n}^{\al},A^{\mu \a}_{n}]=-i \th_{n}({\b}_{T}){\w{A}}_{n}^{\mu}, \ \ \ \  [G_{n}^{\al},B^{\mu \a}_{n}]=-i \th_{n}({\b}_{T}){\w{B}}_{n}^{\mu}, 
$$
$$
[G_{n}^{\al},{\w{A}}^{\mu}_{n}]=-i \th_{n}({\b}_{T}){A}_{n}^{\mu \a}, \ \ \ \  [G_{n}^{\al},{\w{B}}^{\mu}_{n}]=-i \th_{n}({\b}_{T}){B}_{n}^{\mu \a}, 
$$
The thermal vacuum and the thermal creation and annihilation operators are constructed as in the Section 3 by the action of the unitary Bogoliubov mapping 
\be
\left. \left| 0(\b_T ) \right\rangle \! \right\rangle = \prod_{ m > 0} 
e^{iG_m^{\al}}
\left. \left| 0 \right\rangle \! \right\rangle_{\al} ~\prod_{ n > 0} 
e^{iG_n^{\b}} 
\left. \left| 0 \right\rangle \! \right\rangle_{\b}=\prod_{ m > 0}|0(\b_{T})_{m} \ra \ra _{\al} \prod_{ n > 0}|0(\b_{T})_{n} \ra \ra _{\b}  
\label{vacfT}
\ee
It follows that the thermal closed string oscillators have the following vacuum
\be
|0(\b_{T})_{n} \ra \ra = |0(\b_{T})_{n} \ra \ra _{\al}\bigotimes|0(\b_{T})_{n} \ra \ra _{\b}.
\ee
The thermal creation an annihilation operators can be obtained by applying the Bogoliubov mapping to the following pairs of operators at zero temperature  $\{ A, A^{\a},\w{A}, \w{A}^{\a} \}$ and $\{ B, B^{\a},\w{B}, \w{B}^{\a} \}$
\bea
A^{\m}_{n}(\b_T) &= & e^{iG_{n}^{\al}}A ^{\m}_{n}e^{-iG^{\al}_n}=u_{n}(\b_{T}){A}^{\m}_{n}-v_{n}(\b_{T})\w{A}^{\m \a}_{n}, \non
\tilde{A}^{\m}_{n}(\b_T) &=& e^{iG^{\al}_n}\tilde{A} ^{\m}_{n}e^{-iG_{n}^{\al}}=u_{n}(\b_{T})\w{A}^{\m}_{n}-v_{n}(\b_{T}){A}^{\m \a}_{n},
\label{annihT} \\
B^{\m}_{n}(\b_T) &= & e^{iG_{n}^{\b}}B ^{\m}_{n}e^{-iG^{\b}_n}=u_{n}(\b_{T}){B}^{\m}_{n}-v_{n}(\b_{T})\w{B}^{\m \a}_{n}, \non
\tilde{B}^{\m}_{n}(\b_T) &=& e^{iG^{\b}_n}\tilde{B} ^{\m}_{n}e^{-iG_{n}^{\b}}=u_{n}(\b_{T})\w{B}^{\m}_{n}-v_{n}(\b_{T}){A}^{\m \a}_{n},
\label{annihT1}
\eea 
where the coefficients of the linearized Bogoliubov transformations are given by the following relations
\be
u_{n}(\b_{T})=\cosh {\th_{n}}(\b_{T}), \hspace{0.5cm} v_{n}(\b_{T})=\sinh \th_{n}(\b_{T}).
\ee
Since the coordinates of the center of mass of the string 
$\hat{p} , \ \hat{x}$
and the tilde string 
$\hat{\w{p}}, \ \hat{\w{x}}$
commute with the creation and annihilation operators, the Bogoliubov transformation leave them invariant. Then we can define the action of the Bogoliubov transformations on the full string fields $X^{\mu}$ and $\tilde{X}^{\mu}$ and derive the fields of the thermal string as in the case of the open strings.

\subsection{Entropy of D-branes}

An interesting application of the construction presented in the above section is the calculation of the entropy of the thermal closed string in the boundary states associated to the $D$-brane. This is somehow similar to the calculation of the entropy of the thermal open string fields with specific boundary conditions presented before. This calculation is interesting since it contributes to understanding the thermodynamic properties of the $D$-branes.

In order to calculate the entropy, one has to define in a meaningful way a thermal $D$-brane. Such object should exist since the $D$-branes are physical objects in the same way as the strings are. Their microscopic structure at zero temperature is defined by boundary conditions (\ref{bchilbert}) in the closed string channel. The corresponding $D$-brane boundary states were given in the equation (\ref{Dpboundstate}) that takes the following form in the oscillator representation of string excitations
\be
\left| B\right\rangle =N_{p}\delta ^
{\left(d_{\perp }\right) }\left(q^{\mu} - x ^{\mu}\right)
e^{ -\sum\limits_{n=1}^{\infty }A
_{n}^{\dagger\mu }S_{\mu \nu }B_{n}^{\dagger\nu }}\left| 
0\right\rangle.
\label{soltempzero}
\ee
Here, we have denoted by $q^{\m}$ the operators corresponding to the center of mass of 
the closed string and $S_{\m \nu }$ is the diagonal matrix $S_{\m \nu} = 
\left( \eta_{ab},-\delta_{ij} \right)$.
The thermalization of the closed string described in the previous section requires the doubling of the string degrees of freedom, i. e. the addition of the identical copy of the string. Therefore, one should duplicate the $Dp$-brane boundary conditions, too, and impose them to the tilde-string. Then the boundary conditions on the total string fields are
given by the following relations
\bea
\6_{\t}X^{a}|_{\t = 0}(\b_T)\left.\left| B(\b_T) \right\rangle 
\right\rangle&=& 
\6_{\t}\tilde{X}^{a}|_{\t = 0}(\b_T)\left.\left| B(\b_T) 
\right\rangle\right\rangle
=0, \nonumber\\
X^{i}|_{\t = 0}(\beta_T) \left.\left| B(\b_T) 
\right\rangle\right\rangle &=&
\tilde{X}^{i}|_{\t = 0}(\beta_T) \left.\left| B(\b_T) 
\right\rangle\right\rangle = x^i,
\label{defDbrT}
\eea 
where $a = 1, 2, \ldots, p$ and $i = p+1, \ldots, 24$. Note that the center of mass of the string and the tilde string have the same coordinates. By using the properties of the Bogoliubov transformation defined by the equation 
(\ref{vacfT}) one can show that the thermal $Dp$-brane is described by the following boundary state
\be
\left.\left| B(\b_T) \right\rangle\right\rangle = 
N_p^2\delta^{2\left(d_{\perp }\right) }\left(q-x\right)
e^{-\sum\limits_{n=1}^{\infty }\left[A
_{n}^{\dagger\mu }(\b_T)+\tilde{A}
_{n}^{\dagger\mu }(\b_T)\right]
S_{\mu \nu }\left[B_{n}^{\dagger\nu }(\b_T)
 + \tilde{B}_{n}^{\dagger\nu }(\b_T)\right]}
\left.\left| 0 (\b_T)\right\rangle\right\rangle,
\label{thermDp}
\ee 
where 
\be
\delta^{2\left(d_{\perp }\right) 
}\left(q-x\right)=\delta^{\left(d_{\perp }\right) }\left(q-x\right)
\delta^{\left(d_{\perp }\right) }\left( \tilde{q}-x\right)
\label{rev-deltas}
\ee
The thermal state (\ref{thermDp}) can be obtained either by solving the boundary equations (\ref{defDbrT}) 
explicitly or by applying the unitary and tilde invariant transformations to the $Dp$-brane state of the total string at zero temperature. 

The entropy operator is the sum of operators from left- and right-sectors and it is given by the following relation
\begin{eqnarray}
K &=&\sum\limits_{\mu }\sum\limits_{n}\left[ \left( A_{n}^{\mu \dagger
}A_{n}^{\mu }+B_{n}^{\mu \dagger }B_{n}^{\mu }\right) \log \sinh 
^{2}\theta
_{n}\right. +  \nonumber \\
&&\left. -\left( A_{n}^{\mu }A_{n}^{\mu \dagger }+B_{n}^{\mu 
}B_{n}^{\mu
\dagger }\right) \log \cosh ^{2}\theta _{n}\right]. 
\label{entroop1}
\end{eqnarray}
It has all the properties of the entropy operator discussed in the Section 3. The operator $\tilde{K}$ can be constructed from the tilde-string modes. Direct computations \cite{Vancea:2000gr} show that the value of $K$ in the state 
(\ref{thermDp}) is given by the following relation
\begin{eqnarray}
K_{Dp} & = & \left\langle \! \left\langle 
B\left( \beta _{T}\right) \right| \right.
K\left. \left| B\left( \beta _{T}\right) \right\rangle \! \right\rangle
=48\sum_{m=1}^{\infty }\left[\log \sinh ^{2}\theta _{m}
-{\sinh}^{2}\theta_m \log \tanh^{2}\theta _{m}\right] \nonumber \\
&&+2\prod_{m=1}^{\infty }\prod_{\mu=1}^{24}
\prod_{\nu =1}^{24}\sum_{k=0}^{\infty }{\cosh}^{2}\theta_m
\frac{\left( -\right)^{2k+2}S_{\mu \nu }^{2k+2}}{k!\left( k+1\right) 
!}.
\label{braneentropy}
\end{eqnarray}
The quantity $K_{Dp}$ can be interpreted as the intrinsic entropy of the $Dp$-brane at finite temperature. From it, one can derive the free energy and other thermal properties. However, one should note that $K_{Dp}$ diverges as $T\rightarrow 0$ and goes as $\log(-1)$ for $T\rightarrow \infty$. The divergence of the entropy is a typical property of infinite collections of oscillators which needs to be addressed by proper regularization techniques.

\section{Conclusions}

In this review, we have presented the construction of the thermal bosonic string and $D$-brane at finite temperature in the framework of the Thermo Field Dynamics. Since the free strings consist of infinite number of harmonic oscillators, one can apply the TFD method in a direct way. By interpreting the matrix elements of the entropy operator in the thermal states as the entropy of those states, we showed how to derive the thermal entropy of open string fields with all possible boundary conditions and the intrinsic entropy of the $D$-branes. These entropies are normally divergent and should be renormalized, possibly by applying the methods presented in the references. There we can also find the generalization of the TFD formalism to the supersymmetric strings and their boundary states. Some standard results reviewed here were synthesised previously in \cite{PaniagodeSouza:2002nz,Graca:2007bx}.

\end{document}